\documentclass[sigplan,screen]{acmart}

\AtBeginDocument{%
  \providecommand\BibTeX{{%
    Bib\TeX}}}

\usepackage{amsmath,amsfonts}
\usepackage{algorithm}
\usepackage[noend]{algpseudocode} 
\usepackage{graphicx}
\usepackage{textcomp}
\usepackage{xcolor}
\usepackage{subfigure}
\usepackage{xspace}
\usepackage{tikz}

\usepackage{graphicx}
\usepackage{subcaption} 
\usepackage{amsmath}

\usepackage{makecell}
\usepackage{bm}
\usepackage{balance}
\usepackage{hyperref} 
\usepackage[switch]{lineno} 
\usepackage[symbol]{footmisc}
\usepackage{multirow}
\usepackage{float}

\usepackage[most]{tcolorbox}
\tcbset{
  colback=blue!5!white, 
  colframe=white,       
  boxrule=0pt,          
  arc=1mm,              
  left=1mm, right=1mm, top=1mm, bottom=1mm 
}

\def\BibTeX{{\rm B\kern-.05em{\sc i\kern-.025em b}\kern-.08em
    T\kern-.1667em\lower.7ex\hbox{E}\kern-.125emX}}

\copyrightyear{2026}
\acmYear{2026}
\setcopyright{cc}
\setcctype{by}
\acmConference[PASC '26]{Platform for Advanced Scientific Computing Conference}{June 29-July 01, 2026}{Bern, Switzerland}
\acmBooktitle{Platform for Advanced Scientific Computing Conference (PASC '26), June 29-July 01, 2026, Bern, Switzerland}
\acmDOI{10.1145/3815572.3815743}
\acmISBN{979-8-4007-2734-4/2026/06}

\begin{document}

\title{XOR Bidding and Knapsack Formulations for HPC Network Resource Allocation}

\author{Abrar Hossain}
\affiliation{
  \institution{The University of Toledo}
  \city{Toledo}
  \state{Ohio}
  \country{USA}
}
\email{abrar.hossain@utoledo.edu}

\author{Kishwar Ahmed}
\affiliation{
  \institution{The University of Toledo}
  \city{Toledo}
  \state{Ohio}
  \country{USA}
}
\email{kishwar.ahmed@utoledo.edu}

\begin{abstract}
Modern High Performance Computing (HPC) centers face growing challenges in ingesting large and diverse data streams. These issues often create bottlenecks that limit bandwidth use and delay scientific progress. Traditional static allocation and simple queuing methods are not sufficient. This paper presents a dynamic and value based approach to bandwidth allocation. We formalize the problem by incorporating both network and processing constraints. To address it, we introduce two new auction based mechanisms: the Greedy Value Density Auction, which is fast to compute, and the Vickrey–Clarke–Groves (VCG) Knapsack Auction, which offers strong theoretical guarantees. Both mechanisms rely on user bids that include data needs and scientific value. The goal is to maximize the total value of successful transfers, often referred to as social welfare. Simulation results show that our auction mechanisms significantly outperform First Come First Served (FCFS) baselines. In high load conditions, they reduce average and tail completion delays by over 80\%. Predictability also improves, with the Coefficient of Variation of delay falling by 75 to 85\%. Network stability increases as well, with load volatility (Peak to Average Ratio) dropping by up to 60 to 70\%. This value driven and adaptive strategy helps reduce congestion, improves bandwidth use, and ensures fairer access based on scientific importance. 
\end{abstract}

\maketitle



\section{Introduction}

\begin{figure}[t]
    \centering
    \includegraphics[width=0.7\linewidth]{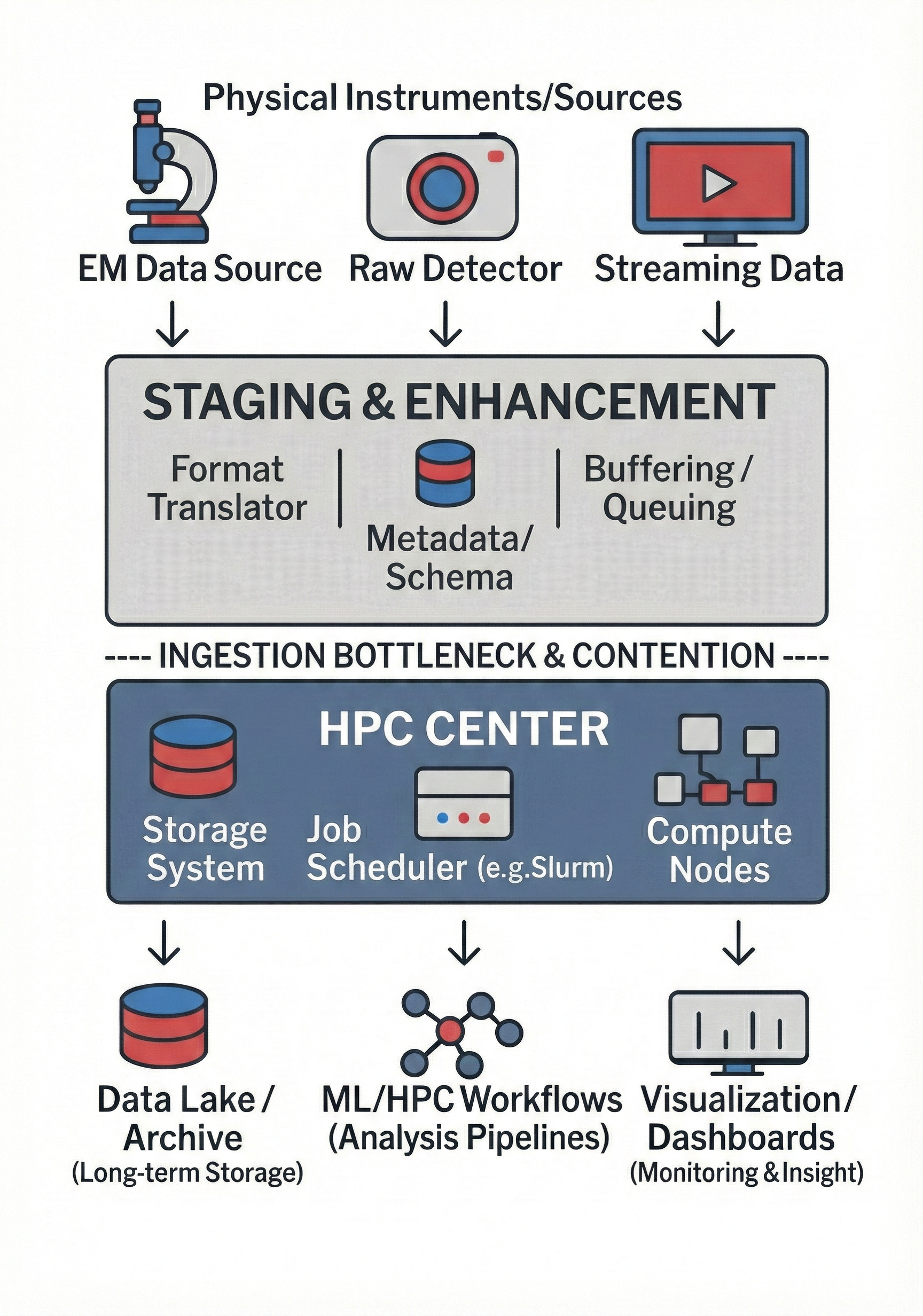}
    \caption{The scientific data pipeline, highlighting heterogeneous sources feeding into a common ingestion interface.}
    \label{fig:pipeline_diagram}
\end{figure}

\textbf{Motivation:} Modern high-performance computing (HPC) centers increasingly function as ``super-facilities,'' integrating large-scale compute resources with diverse external data sources, including advanced scientific instruments like light sources and electron microscopes, as well as distributed sensor networks~\cite{Dart2021, Chard2016}. Consequently, reliable and efficient data movement across potentially wide-area networks has become as critical to scientific productivity as computation itself. As outlined in Figure ~\ref{fig:pipeline_diagram}, the primary challenge lies in ingesting massive, heterogeneous data streams originating from these sources, which exhibit wide variations in communication protocols (e.g., file transfer vs. streaming), data formats, metadata schemas, generation rates (spanning orders of magnitude), and reliability characteristics \cite{Monti2011}. This inherent diversity often necessitates custom interfaces or gateways, complicates standardization efforts~\cite{Enders2024}, and leads to unpredictable, bursty load patterns at the HPC edge, challenging traditional infrastructure assumptions~\cite{Giannakou2024}. As data traverses staging systems towards core HPC storage and processing, it frequently encounters a significant bandwidth bottleneck at the ingestion point, affecting not just network links but potentially intermediate buffers and file systems~\cite{Benoit2024}. Uncoordinated demands from disparate users and instruments exacerbate this bottleneck, resulting in network congestion, inefficient resource utilization characterized by oscillations between overload and underuse \cite{Giannakou2024}, and significant transfer delays that directly impede end-to-end scientific workflows and leave expensive compute resources idle~\cite{Ding2023, Welborn2025}. Effectively managing access to this shared, limited ingestion bandwidth is therefore paramount for enabling timely, large-scale scientific analysis.

\textbf{Where Existing Data Movement Tools Fall Short:} While various tools---such as SCP/SFTP\cite{kolano2015automatically}, Rsync\cite{uazRsync}, GridFTP \cite{gctGridFTP}, and proprietary UDP-based solutions like Aspera\cite{ibmAspera}---are commonly employed for data transfers, they often exhibit significant limitations when faced with the scale, heterogeneity, and dynamic nature of modern scientific workflows. Fundamentally, these point-to-point tools lack the adaptability required for dynamic end-to-end path conditions and are not inherently fault-tolerant in less predictable, highly congested environments \cite{Monti2011}. As detailed further in Section \ref{subsec:limitations_tools}, the inability of these tools to optimally arbitrate mixed-priority workloads at the edge motivates the need for a more intelligent, value-aware data movement framework.

\textbf{Our Contributions:}
To address the critical challenges of managing constrained ingestion bandwidth amidst heterogeneous and dynamic demands, this paper makes the following contributions:
\begin{itemize}
    \item We formalize the HPC ingestion bandwidth allocation problem considering heterogeneous demands and priorities.
    \item We propose two novel, value-aware auction mechanisms: \textit{Greedy Value-Density} and \textit{VCG Knapsack} for HPC data ingestion.
    \item We detail the value-driven allocation and payment logic specific to each proposed mechanism.
    \item We analyze the trade-offs between the VCG mechanism's truthfulness and the Greedy mechanism's efficiency.
    \item We demonstrate through simulation the potential of auction-based strategies to mitigate ingestion bottlenecks and improve resource utilization compared to traditional methods.
\end{itemize}
\section{Background}
\label{sec:background}

\begin{figure*}[t!]
\centering
\subfigure[Hourly capacity demand at ALCF often exceeds 90\% utilization, indicating frequent network congestion.]{\label{fig:capacity_utilization_plot} 
\includegraphics[width=0.30\textwidth]{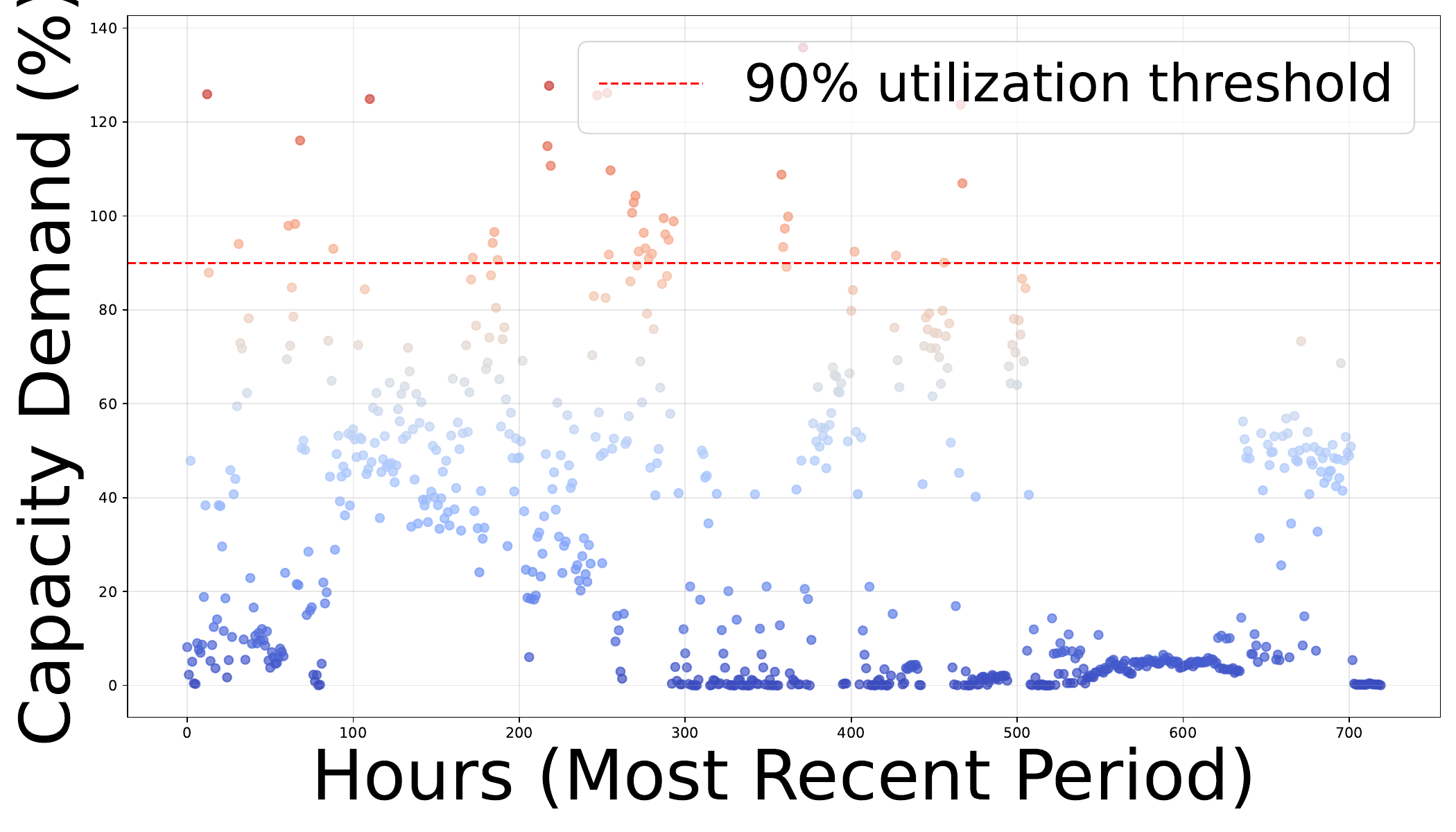}}\hspace{0.02\textwidth}
\subfigure[Evolution of average transfer size (GB) and total number of transfers per year at ALCF (2014-2024).]{\label{fig:yearly_transfer_stats_plot} 
\includegraphics[width=0.30\textwidth]{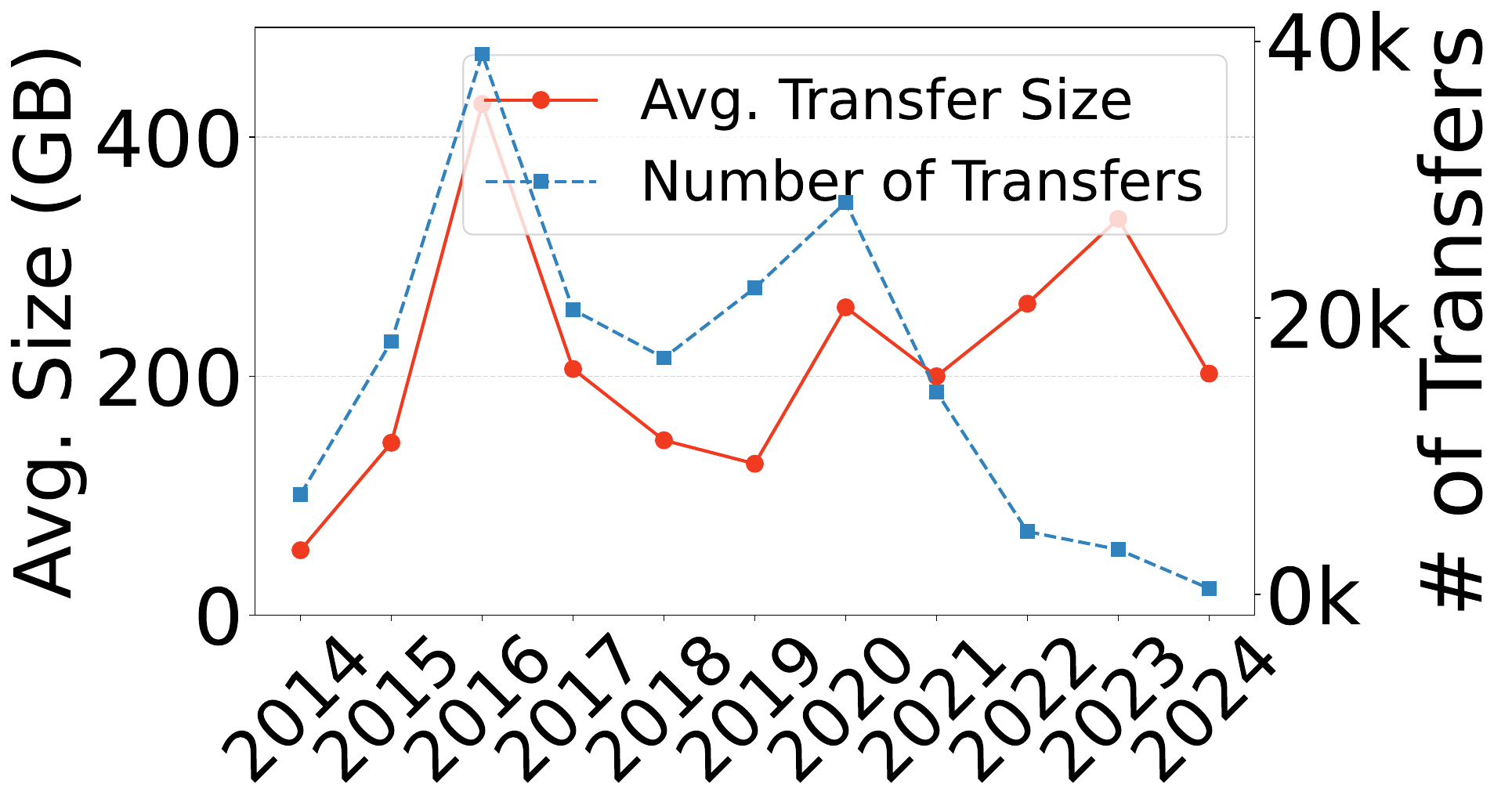}}\hspace{0.02\textwidth}
\subfigure[Yearly distribution of GridFTP transfer sizes (GB) at ALCF, illustrating increasing scale and persistent variability.]{\label{fig:transfer_size_boxplot_plot} 
\includegraphics[width=0.30\textwidth]{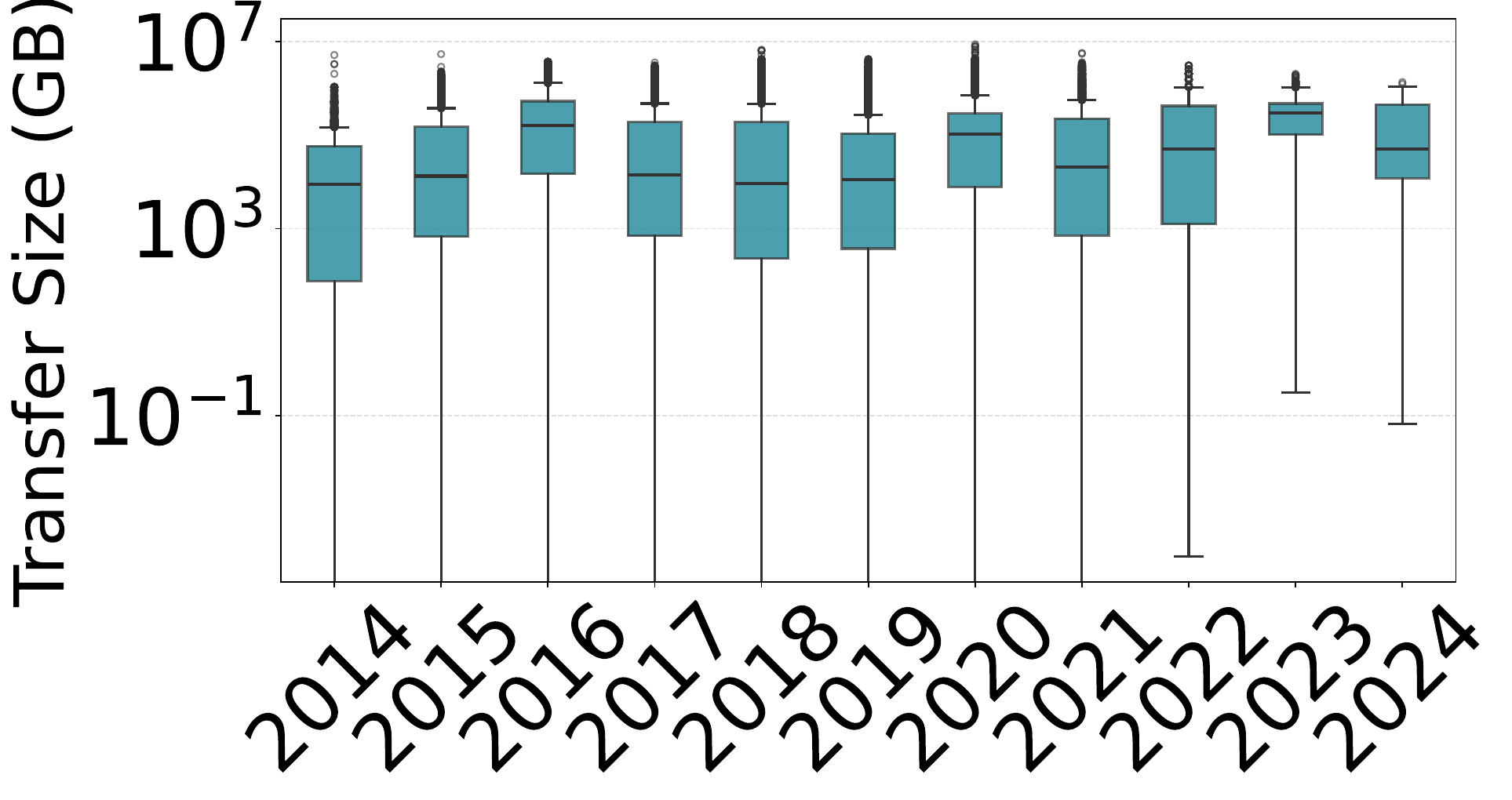}}
\vspace{-3mm}
\caption{Observed data ingestion characteristics at ALCF \cite{alcfGridFTP}.}
\label{fig:figset1_network_workload_trends} 
\end{figure*}

HPC systems are increasingly burdened by data ingestion, where data transfer and I/O constraints critically limit end-to-end scientific workflows. Analysis of extensive operational data from large-scale facilities like the Argonne Leadership Computing Facility (ALCF)~\cite{alcfGridFTP} reveals persistent challenges in managing data ingress efficiently. These challenges, stemming from workload characteristics and user behaviors, motivate the need for improved ingestion strategies.

\subsection{Bottlenecks and Inefficiencies}
\label{subsec:bottlenecks}

Symptoms of ingestion bottlenecks include significant network saturation and unpredictable latency spikes, frequently arising from competing data flows that degrade each other's throughput, even across high-speed research networks~\cite{Giannakou2024}. The volatility of network demand is clearly illustrated in Figure ~\ref{fig:capacity_utilization_plot}, which shows hourly GridFTP data volume at ALCF, normalized against a representative operational pathway capacity; demand frequently spikes above a 90\% threshold, indicating periods of intense contention for this shared resource. Current infrastructure often struggles to handle such competing transfers efficiently~\cite{Giannakou2024}, leading to substantial delays in data availability and potentially underutilized compute resources awaiting necessary input~\cite{Benoit2024}. This contention translates to significant and unpredictable transfer latencies; for instance, traditional methods in microscopy workflows incurred waits of 10-20 minutes, a delay drastically reduced by optimized streaming approaches~\cite{Welborn2025}. Manual or semi-automated ``store-and-forward'' processes further contribute to these inefficiencies and can even lead to data loss~\cite{Welborn2025}.

The nature of the workload itself, as characterized by the ALCF data, contributes to these inefficiencies. As can be seen in Figure ~\ref{fig:yearly_transfer_stats_plot}, while the number of transfers has fluctuated, the average transfer size has shown a significant increasing trend, particularly in recent years, placing concentrated pressure on the ingest infrastructure. The wide range of transfer sizes, spanning orders of magnitude from megabytes to terabytes, is evident year-over-year (Figure ~\ref{fig:transfer_size_boxplot_plot} and Figure ~\ref{fig:cumulative_transfer_size_dist}), demanding a system that can accommodate diverse data scales. Furthermore, uncoordinated user-initiated transfers create bursty network patterns, causing transient congestion during peak times~\cite{Giannakou2024}, while leaving bandwidth underutilized at other times. This is compounded by heterogeneous user behaviors: a small fraction of users often contributes a disproportionate amount to peak traffic (Figure ~\ref{fig:user_peak_pareto}), and users exhibit diverse predictability in their transfer patterns (Figure ~\ref{fig:user_predictability_distribution}) and varying data interaction modes, such as differing PUT versus GET size characteristics (Figure ~\ref{fig:user_get_vs_put}) or a wide range in the number and average size of their transfers (Figure ~\ref{fig:user_count_vs_size}). The Gini coefficient tracking monthly transfer volume (Figure ~\ref{fig:gini_coefficient_time}) further illustrates the fluctuating concentration of demand. This dynamic makes workflows externally bandwidth-bound~\cite{Ding2023} and can overwhelm intermediate storage or burst buffers, potentially forcing experiments to slow down or causing data loss in high-rate streaming scenarios.

\begin{tcolorbox}[colback=cyan!5!white, sharp corners]
\textbf{Takeaway 1:} \textit{Observed HPC data ingestion, as illustrated by ALCF data (Figs.~\ref{fig:figset1_network_workload_trends}-\ref{fig:figset3_detailed_transfer_patterns}), is characterized by volatile network demand, evolving workload profiles with increasingly large transfers, and significant heterogeneity in data sizes and user behaviors, collectively leading to frequent performance bottlenecks and resource inefficiencies.}
\end{tcolorbox}

\subsection{Limitations of Existing Data Ingestion Strategies}
\label{subsec:limitations_tools}

\begin{figure*}[t!]
\centering
\subfigure[Cumulative traffic contribution, showing a small fraction of users responsible for a large percentage of peak demand.]{\label{fig:user_peak_pareto} 
\includegraphics[width=0.30\textwidth]{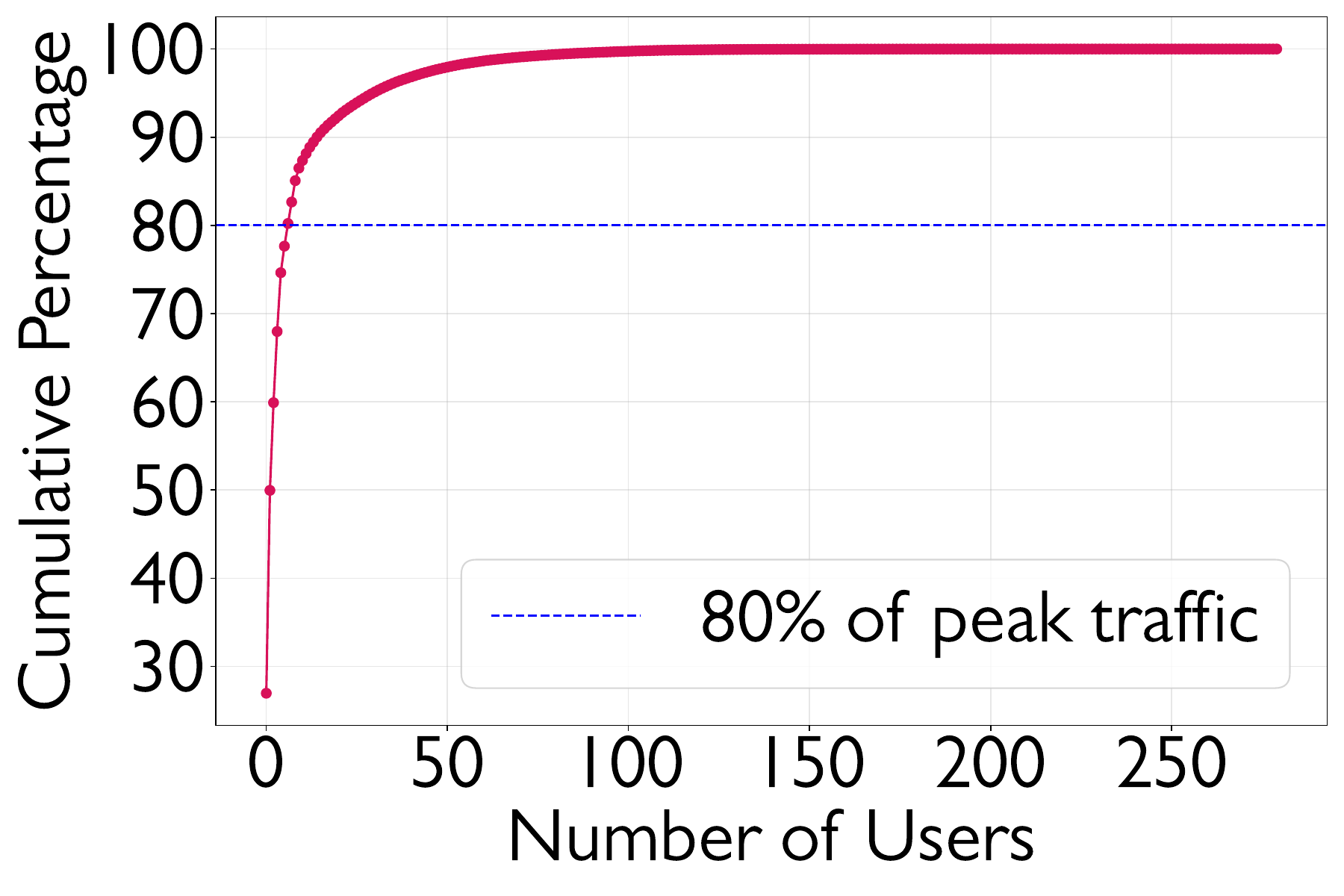}}\hspace{0.02\textwidth}
\subfigure[Distribution of user transfer predictability scores, highlighting diverse and often unpredictable user behavior.]{\label{fig:user_predictability_distribution} 
\includegraphics[width=0.30\textwidth]{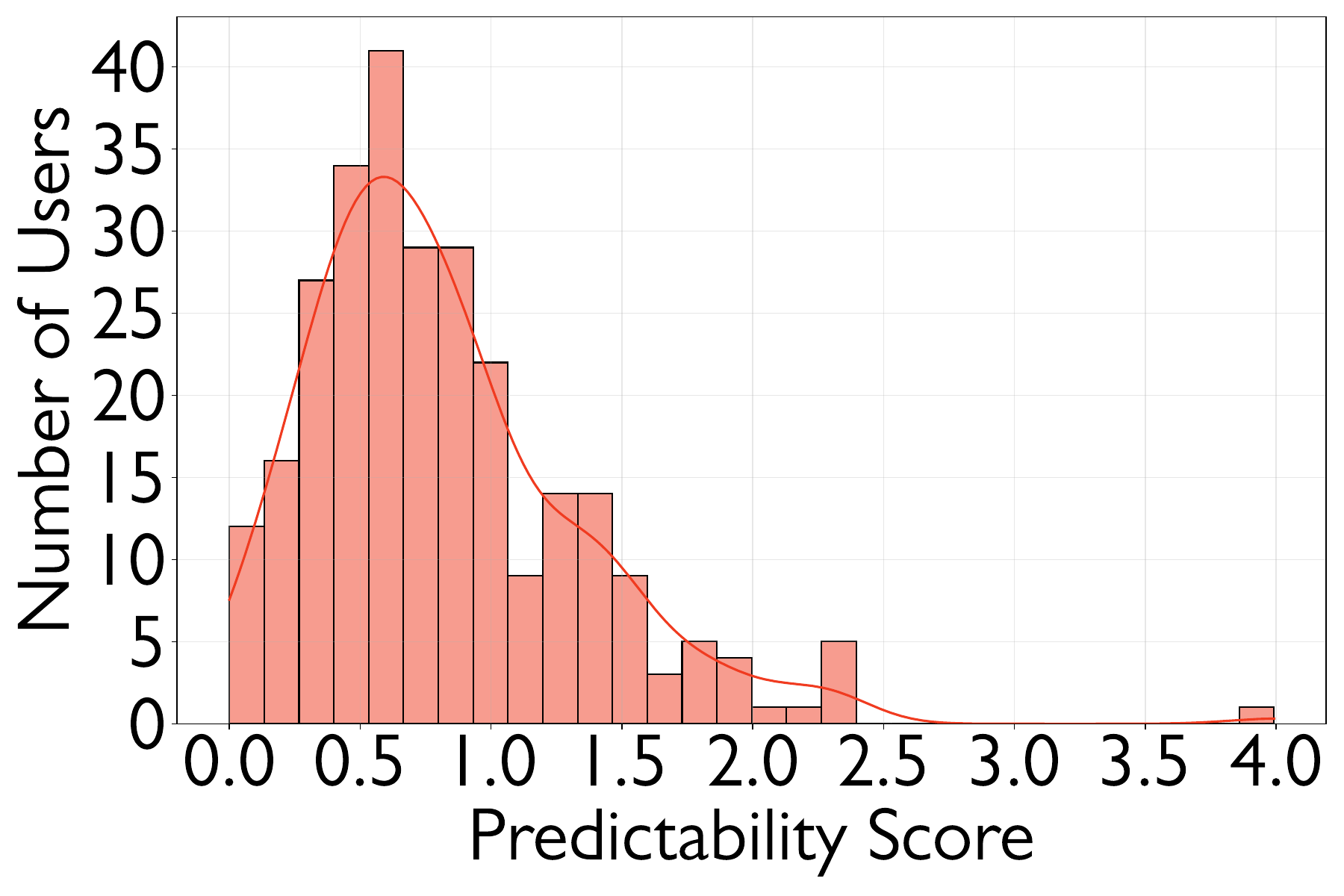}}\hspace{0.02\textwidth}
\subfigure[Gini coefficient of monthly transfer volume over time, illustrating fluctuations in demand concentration among users.]{\label{fig:gini_coefficient_time} 
\includegraphics[width=0.30\textwidth]{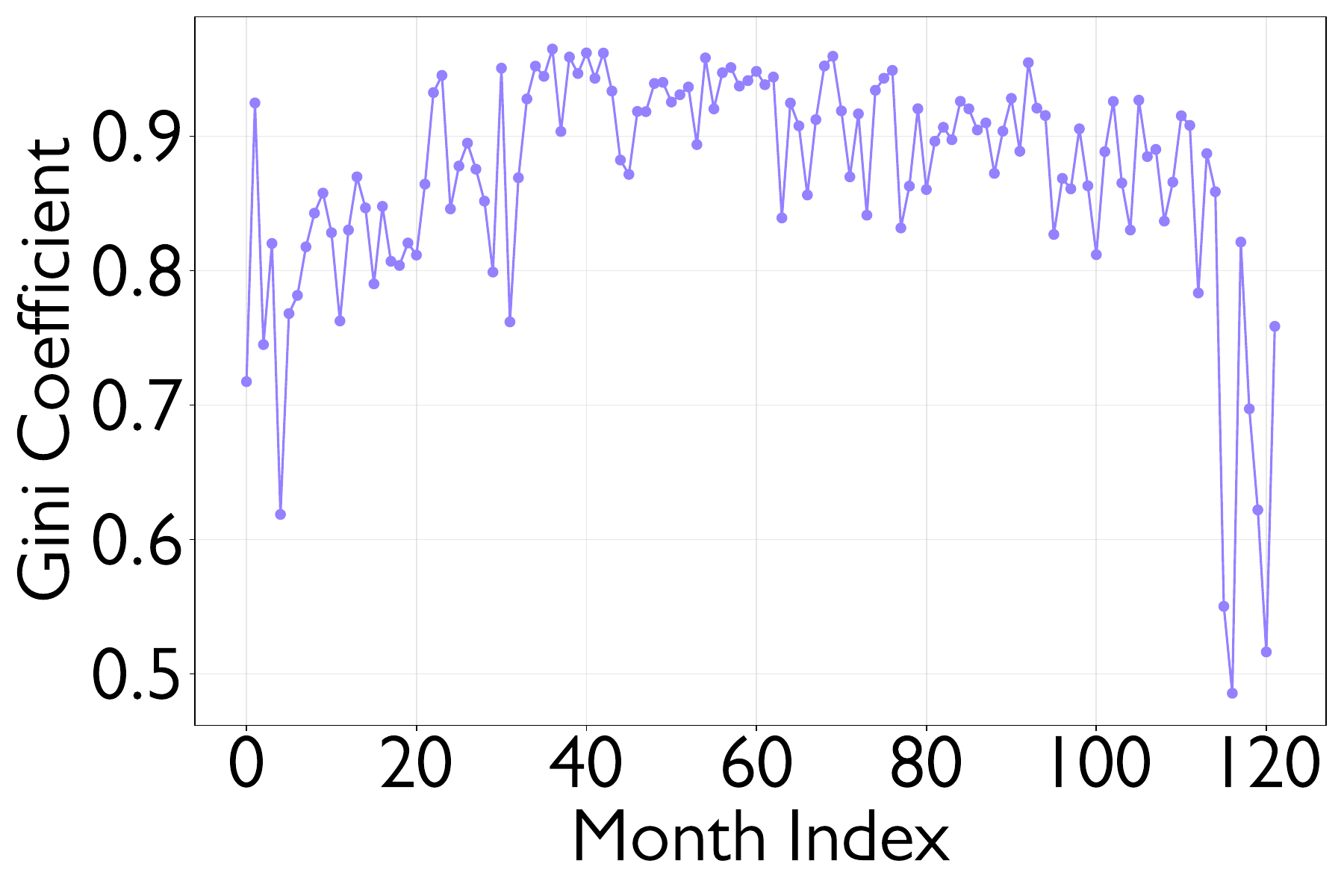}}
\vspace{-3mm}
\caption{Analysis of user behavior characteristics from ALCF GridFTP data~\cite{alcfGridFTP}.}
\label{fig:figset2_user_behavior_analysis} 
\end{figure*}

Existing tools for transferring data into HPC centers exhibit significant limitations when confronted with the scale and heterogeneity of modern scientific workflows, as evidenced by operational data (Figs.~\ref{fig:figset1_network_workload_trends},~\ref{fig:figset2_user_behavior_analysis}, and~\ref{fig:figset3_detailed_transfer_patterns}). 

\begin{itemize}
    \item \textbf{SCP/SFTP and Rsync:} Ubiquitous tools like SCP/SFTP\cite{kolano2015automatically}, though secure, are constrained by single-stream transfer and high encryption overhead, rendering them inefficient for bulk data movement over high-latency networks~\cite{Arslan2018, Chard2016}. Rsync\cite{uazRsync} faces similar performance bottlenecks during initial transfers of large datasets or numerous small files due to computational overhead and latency sensitivity.
    
    \item \textbf{Globus/GridFTP:} As the de-facto standard for high-performance, reliable transfers between well-provisioned sites~\cite{gctGridFTP, Chard2016, Dart2021}, GridFTP handles terabyte datasets well but requires capable Data Transfer Nodes (DTNs) at both ends---a requirement often unmet by diverse scientific instruments or edge devices. Furthermore, its FIFO-based scheduling struggles to optimally manage resources under mixed-priority workloads or dynamic congestion, given observed workload variability (Figs.~\ref{fig:yearly_transfer_stats_plot},~\ref{fig:transfer_size_boxplot_plot}, and~\ref{fig:cumulative_transfer_size_dist}). While it historically required enhancements for extreme-scale single files~\cite{petrie2020coordinating}, scripting directly with standalone GridFTP can yield high performance, but it demands expert tuning across varying conditions and lacks user-friendly failure recovery~\cite{Arslan2018}. Additionally, NERSC observations suggest endpoint complexity can limit its adoption for certain data collection scenarios~\cite{Giannakou2024}.
    
    \item \textbf{Proprietary UDP-based Tools:} Solutions like Aspera\cite{ibmAspera} excel on challenging networks by overcoming TCP limitations, but they introduce deployment barriers through licensing requirements and endpoint software needs. Furthermore, their aggressive bandwidth consumption can negatively impact fairness in shared environments.
\end{itemize}

Consequently, no single existing tool comprehensively addresses the multifaceted challenges of heterogeneous data sources, variable network conditions, and dynamic user demands in HPC data ingestion~\cite{Monti2011}.

\begin{tcolorbox}[colback=cyan!5!white, sharp corners]
\textbf{Takeaway 2:} \textit{Current data transfer tools, despite individual merits, possess inherent limitations in performance, scalability, usability, or deployment flexibility that hinder their effectiveness when confronted with the complex, dynamic, and heterogeneous data ingestion environment observed in production HPC systems (Figs.~\ref{fig:figset1_network_workload_trends}-~\ref{fig:figset3_detailed_transfer_patterns}).}
\end{tcolorbox}

\subsection{Need for Adaptive Ingestion Bandwidth Mechanism}
\label{subsec:need_adaptive}

\begin{figure*}[t!]
\centering
\subfigure[Cumulative distribution of transfer sizes (GB) per year, showing shifts towards larger transfers dominating total volume.]{\label{fig:cumulative_transfer_size_dist} 
\includegraphics[width=0.30\textwidth]{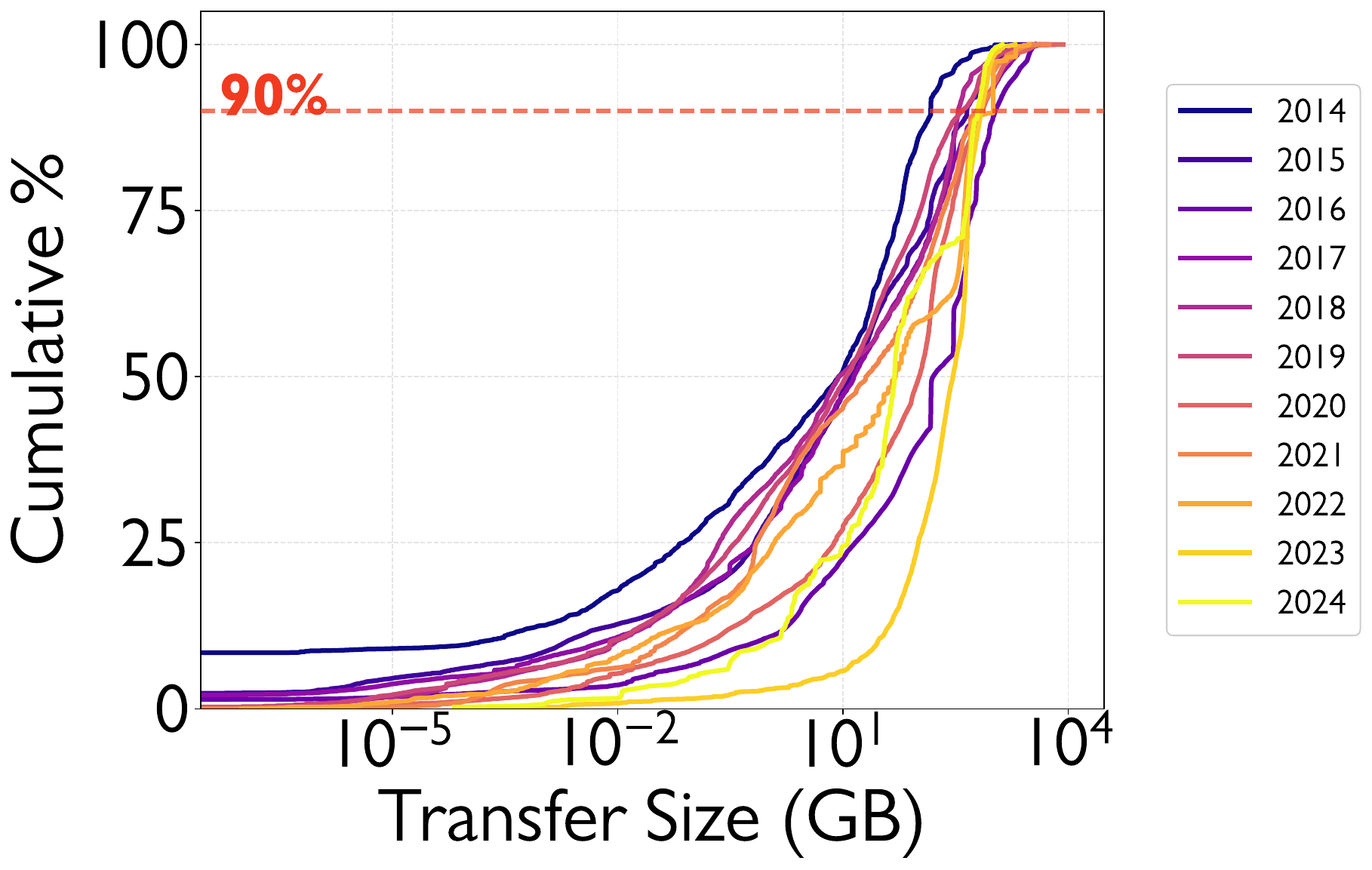}}\hspace{0.02\textwidth}
\subfigure[Relationship between total PUT (upload) and GET (download) transfer sizes (GB) per user.]{\label{fig:user_get_vs_put} 
\includegraphics[width=0.30\textwidth]{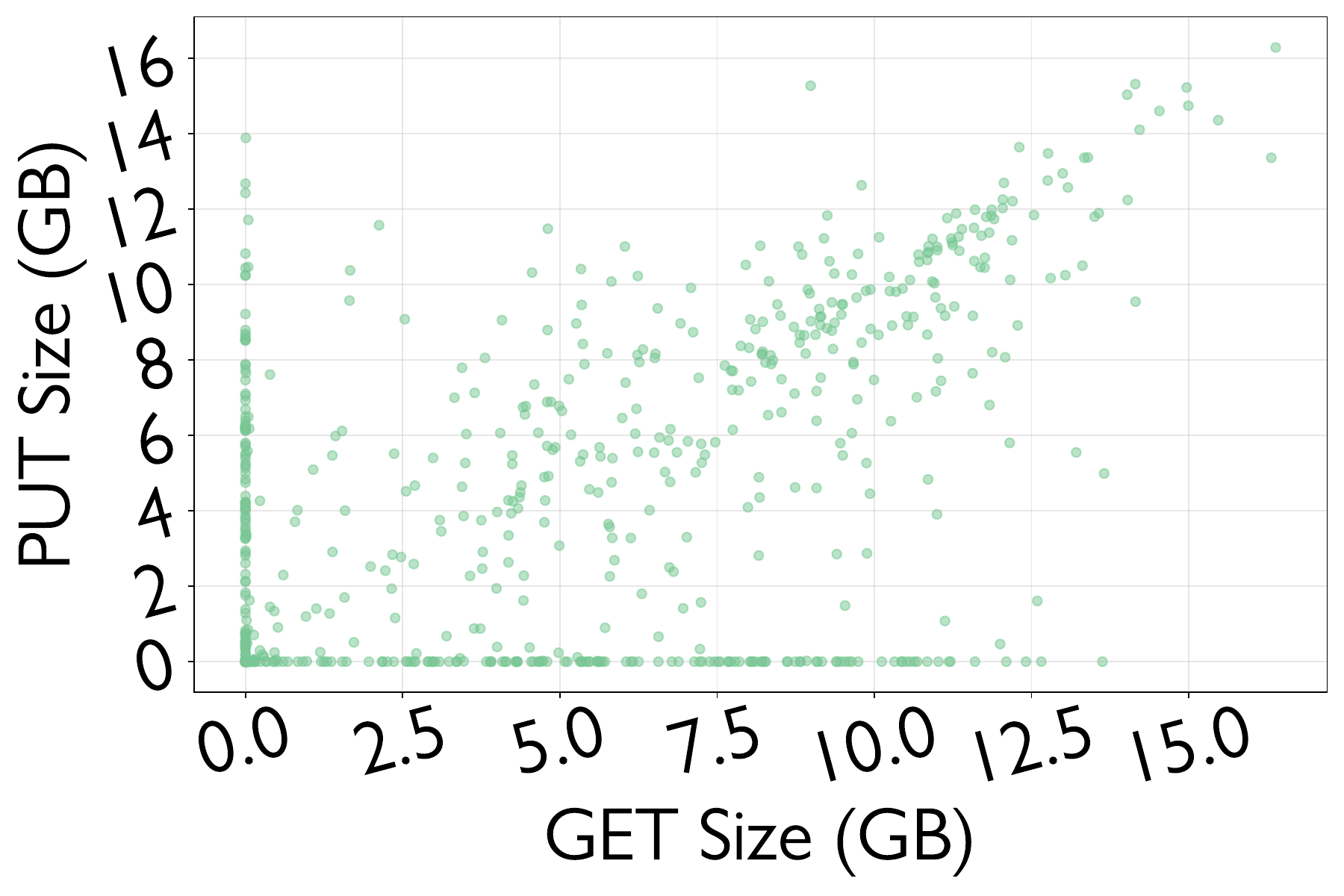}}\hspace{0.02\textwidth}
\subfigure[Mean transfer size (GB) versus total number of transfers per user, highlighting heterogeneity.]{\label{fig:user_count_vs_size} 
\includegraphics[width=0.30\textwidth]{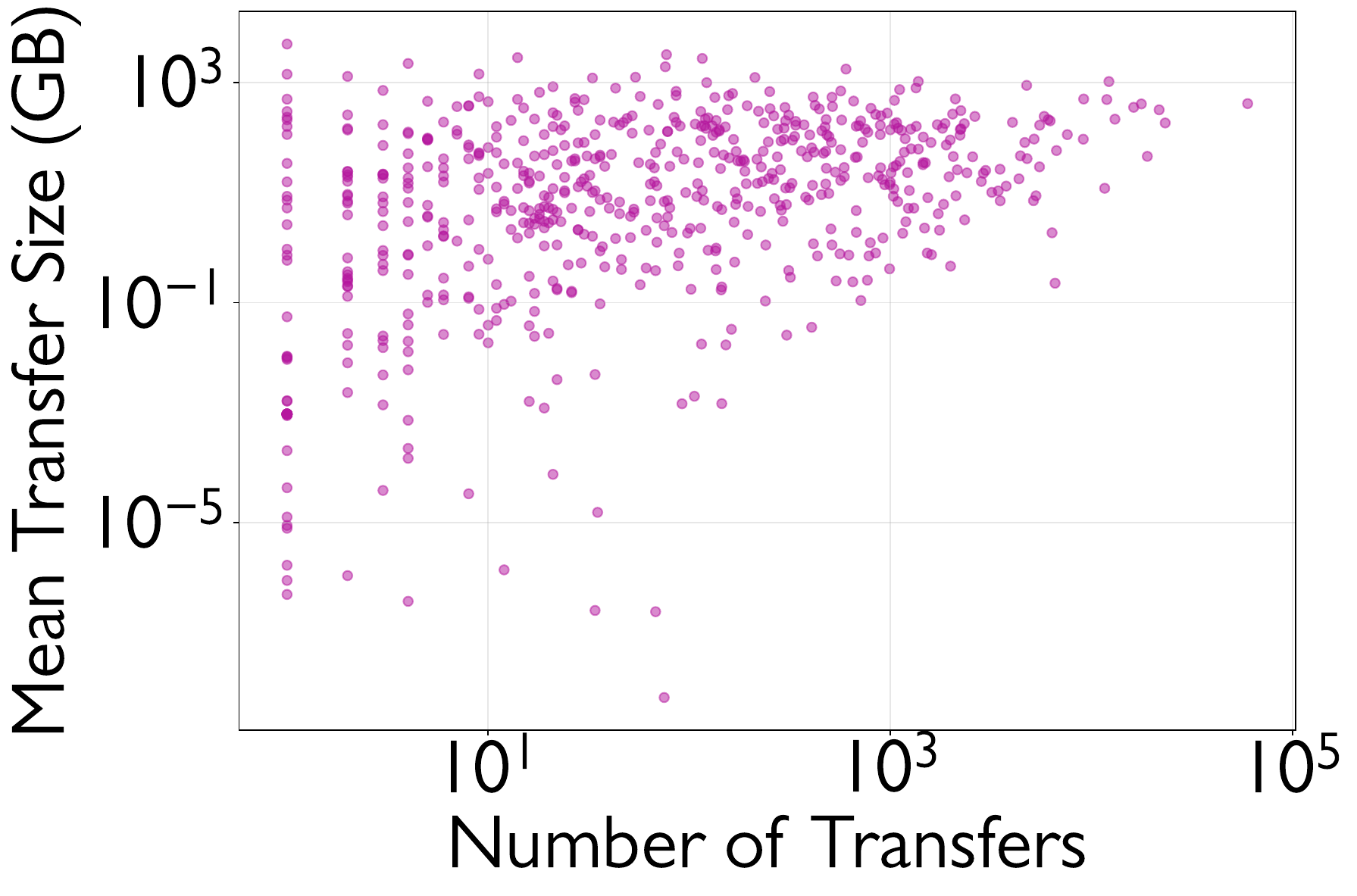}}
\vspace{-3mm}
\caption{Detailed characterization of data transfer patterns from ALCF GridFTP logs~\cite{alcfGridFTP}.}
\label{fig:figset3_detailed_transfer_patterns} 
\end{figure*}

The combination of escalating data volumes and transfer sizes (Figs.~\ref{fig:yearly_transfer_stats_plot},~\ref{fig:transfer_size_boxplot_plot},~\ref{fig:cumulative_transfer_size_dist}), inherent heterogeneity across data sources and user behaviors (Figs.~\ref{fig:user_peak_pareto},~\ref{fig:user_predictability_distribution},~\ref{fig:gini_coefficient_time},~\ref{fig:user_get_vs_put},~\ref{fig:user_count_vs_size}), coupled with the identified limitations of current transfer tools and static scheduling strategies, underscores the necessity for a new approach to data ingestion management within HPC environments. This confluence of factors, leading to volatile demand and frequent congestion (Figure ~\ref{fig:capacity_utilization_plot}), demands a shift towards more intelligent, adaptive, and scalable ingestion frameworks.  Essential capabilities for such frameworks include the ability to dynamically adjust to variations in source characteristics, data properties, and real-time network conditions. They must be designed to efficiently scale with numerous concurrent data streams originating from diverse sources. At the same time, they should continuously optimize the use of critical resources particularly network bandwidth at HPC ingress points to reduce congestion and improve overall throughput. Furthermore, ensuring fairness in resource allocation among competing scientific workflows and users with disparate requirements is crucial. Automating the orchestration of these complex data movements is also vital to reduce operational burden. Addressing these multifaceted requirements with advanced ingestion frameworks is essential to easing current bottlenecks and fully harnessing the integration of diverse, large-scale data sources with HPC capabilities for accelerated scientific discovery.

\begin{tcolorbox}[colback=cyan!5!white, sharp corners]
\textbf{Takeaway 3:} \textit{The observed characteristics of HPC data ingestion (Figs.~\ref{fig:figset1_network_workload_trends}-\ref{fig:figset3_detailed_transfer_patterns}) and the limitations of current tools necessitate adaptive, resource-aware frameworks to optimize performance and utilization at the HPC-edge.}
\end{tcolorbox}

\section{Core Concepts}
\label{sec:preliminaries}

We briefly introduce core auction concepts pertinent to our bandwidth allocation mechanisms.

\textbf{Auction Mechanism.} A process for allocating a scarce resource (here, HPC network bandwidth) among competing agents (users or experiments) based on submitted bids\cite{krishna2009auction}. It comprises rules for bidding, selecting winners (\textit{Allocation Rule}), and determining payments (\textit{Payment Rule}).

\textbf{Bid.} A request submitted by an agent, specifying their bandwidth requirement (derived from file size and slot duration) and a declared value for obtaining that bandwidth within the time slot.

\textbf{Valuation ($V_j$).} An agent's true, private utility or benefit derived from successfully completing transfer $j$. This may differ from their declared bid value.

\textbf{Utility ($u_j$).} An agent’s utility is the net benefit they receive from participating in the system. If an agent wins the auction and pays a price, their utility is the difference between the value they place on the item and the price they pay. If they do not win, their utility is zero. Rational agents act in ways that aim to maximize this utility.

\textbf{VCG (Vickrey-Clarke-Groves) Mechanism.} A class of auction mechanisms known for being truthful and (typically) allocatively efficient\cite{caminati2015vickrey}. VCG payments are based on the opportunity cost (or externality) a winner imposes on others by consuming resources.

\textbf{Value Density.} A metric used in greedy allocation, defined as the bid value divided by the required resource amount (e.g., value per unit of bandwidth). The Greedy mechanism prioritizes bids with higher value density.

\textbf{Truthfulness (Incentive Compatibility).} A desirable property where an agent's optimal strategy is always to bid their true valuation ($V_j$). VCG mechanisms are designed to be truthful.

\textbf{Social Welfare (SW)\cite{cornellCS6840}.} The sum of the true valuations of the winning bidders:
\[
SW = \sum_{j \in \text{Winners}} V_j.
\]
A key goal is \textit{Allocative Efficiency}, where the allocation rule maximizes social welfare. Mechanisms typically maximize the sum of reported bid values, relying on truthfulness to align this with maximizing true $SW$.

\section{Problem Formulation}

We address allocating limited data ingestion resources in a high-performance computing (HPC) environment. The system has finite network ingestion bandwidth $W$ (e.g., GB/s) and potentially separate downstream processing capacity $P$. These resources are shared among $K$ scientific experiments competing for access during a specific allocation period. Each experiment $k \in \{1, 2, \dots, K\}$ submits multiple bids. The $j$-th bid from experiment $k$ is $t_k^j = \{r_k^j, b_k^j\}$, where $r_k^j$ is the requested ingestion rate and $b_k^j$ is the bid valuation. Let $J_k$ be the number of bids from experiment $k$. We distinguish the submitted bid $b_k^j$ from the experiment's true, private valuation $V_k^j$ for rate $r_k^j$. A central aim is designing a mechanism that encourages truthful bidding: $b_k^j = V_k^j$.

Let $x_k^j$ be a binary decision variable indicating whether experiment $k$'s $j$-th bid wins:
\[
x_k^j = \begin{cases}
1, & \text{if experiment } k \text{ wins bid } j, \\
0, & \text{otherwise.}
\end{cases}
\]
We adopt XOR bidding: each experiment wins at most one bid, reflecting scenarios where experiments have alternative plans but need only one fulfilled. For each experiment $k$:

\begin{equation}
\sum_{j=1}^{J_k} x_k^j \leq 1, \quad \forall k \in \{1, 2, \dots, K\}
\label{eq:exclusive_allocation}
\end{equation}

The allocation must respect physical limitations. Total allocated bandwidth cannot exceed available capacity, and may be limited by processing capacity $P$:

\begin{equation}
\sum_{k=1}^{K} \sum_{j=1}^{J_k} r_k^j x_k^j \leq \min(W, P)
\label{eq:capacity_constraint}
\end{equation}

If experiment $k$'s $j$-th bid wins ($x_k^j = 1$), it pays $p_k^j$. Its net utility based on true valuation is:
\[
u_k^j = \begin{cases}
V_k^j - p_k^j, & \text{if } x_k^j = 1, \\
0, & \text{otherwise.}
\end{cases}
\]

Assuming truthful bidding ($b_k^j = V_k^j$), the system maximizes social welfare—the sum of winning bid valuations:
\[
SW = \sum_{k=1}^{K} \sum_{j=1}^{J_k} V_k^j x_k^j.
\]
The Winner Determination Problem (WDP) finds allocation $x = \{x_k^j\}$ maximizing reported social welfare subject to XOR and capacity constraints:
\begin{equation}
\begin{aligned}
\text{maximize} \quad & \sum_{k=1}^{K} \sum_{j=1}^{J_k} b_k^j x_k^j \\
\text{subject to} \quad 
& \text{Equation: } \eqref{eq:exclusive_allocation} \text{ and } \eqref{eq:capacity_constraint} \\
& x_k^j \in \{0,1\}, \quad \forall k \in \{1, 2, \dots, K\}, j \in \{1, 2, \dots, J_k\}
\end{aligned}
\label{eq:auction_objective}
\end{equation}

We assume submitted valuations $b_k^j$ reflect genuine scientific benefit $V_k^j$ from obtaining bandwidth allocation $r_k^j$. We solve this WDP and determine payments $p_k^j$ using mechanisms outlined in the next section.
\section{Proposed Allocation Mechanisms}
\label{sec:proposed_mechanisms}

\begin{figure}
    \centering
    \includegraphics[width=0.9\linewidth]{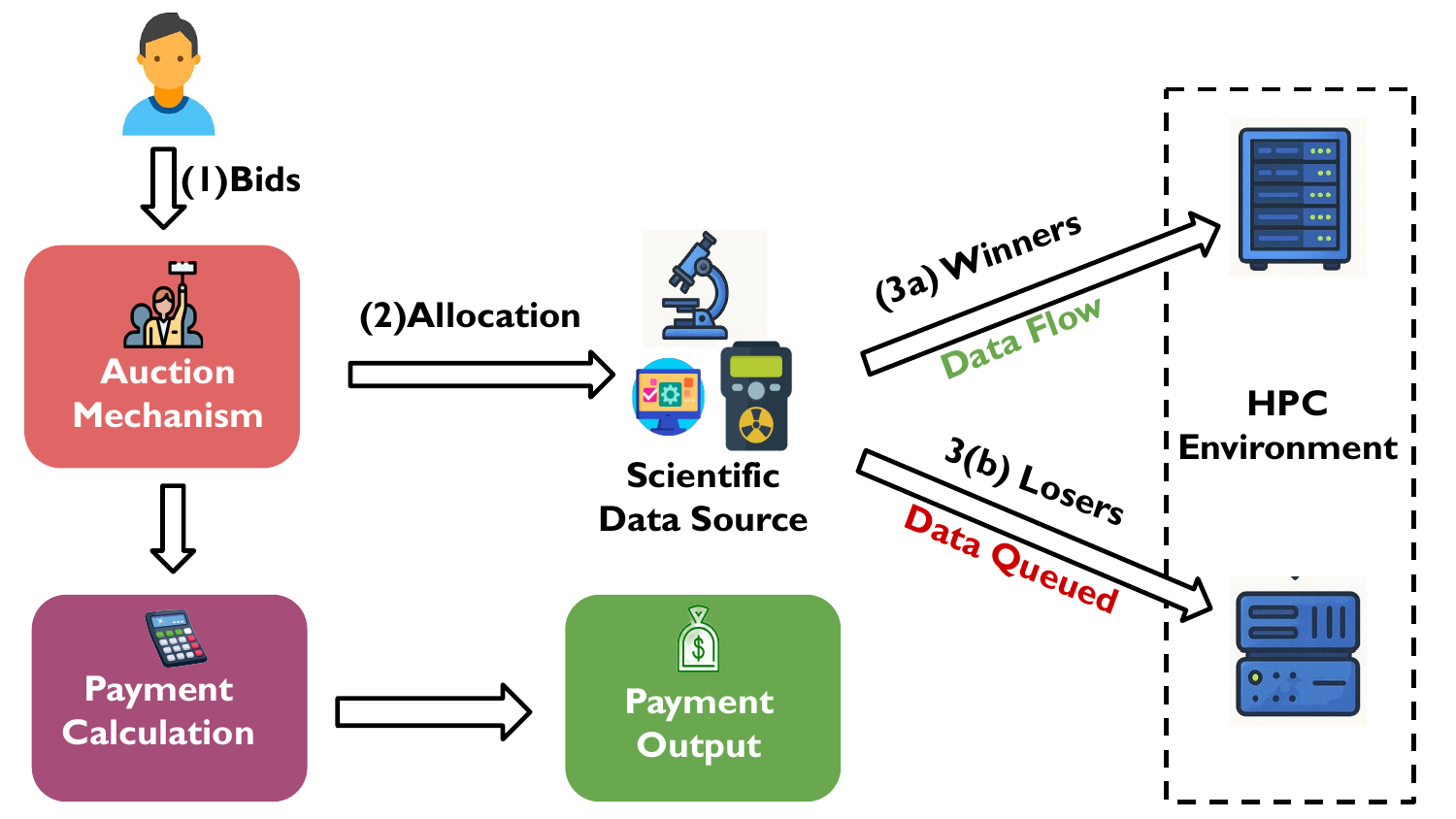}
    \caption{Overview of Auction Based Bandwidth Allocation }
    \label{fig:block_diagram}
\end{figure}

To address network congestion and enable fair allocation of shared HPC bandwidth $B$, we propose two auction-based mechanisms executed periodically (e.g., hourly) to determine bandwidth allocation for the subsequent time slot. A brief overview is shown in Figure  \ref{fig:block_diagram}.

Each user $i$ submits a bid $\text{bid}_i = (\text{request\_details}_i, b_i)$, where $\text{request\_details}_i$ includes the user identifier and file size ($\text{file\_size}_i$), and $b_i$ represents their willingness-to-pay for completing the transfer. The required bandwidth is $\text{bw\_req}_i = \text{file\_size}_i / \text{slot\_duration}$. The auction mechanism selects a winning set $W$ such that $\sum_{i \in W} \text{bw\_req}_i \leq B$ while optimizing the allocation objective.

The first mechanism is a \textbf{Greedy Value Density Auction} (Algorithm \ref{alg:greedy_auction_compact}), which prioritizes bids based on their value density for computational efficiency. The mechanism calculates required bandwidth $c_i = s_i / T_{sec}$ for each bid and filters out bids with non-positive values or those exceeding capacity $C$. For eligible bids, it computes value density $d_i = v_i / c_i$ (value per unit bandwidth) and sorts them in descending order. The allocation proceeds by iterating through the sorted list, adding bids to the winning set $W_{temp}$ while capacity $C_{alloc}$ allows. The critical density $d^*$ is defined as the value density of the highest-density bid that could not be included (or zero if all bids accepted). Each winner $j$ pays $p_j = d^* \times c_j$, meaning winners pay a price per unit bandwidth determined by the most competitive losing bid.

The second mechanism is a \textbf{VCG Auction} based on the Vickrey-Clarke-Groves principle (Algorithm \ref{alg:vcg_knapsack_compact}), which incentivizes truthful bidding and maximizes total social welfare. The allocation is formulated as a 0/1 Knapsack problem. The continuous capacity $C$ and bandwidth requirements $c_i$ are converted to integers using scaling factor $\alpha$: $C' = \lfloor C \times \alpha \rceil$ and $w_i' = \lfloor c_i \times \alpha \rceil$. The Knapsack problem is solved via dynamic programming to find winners $W_{ids}$ that maximize social welfare $SW_{all} = \sum_{i \in W_{ids}} v_i$. For each winner $j$, payment is calculated as $p_j = \max(0, SW_{-j} - (SW_{all} - v_j))$, where $SW_{-j}$ is the optimal welfare without bidder $j$. This ensures winners pay for the externality their inclusion imposes on other bidders. Both mechanisms output winning bids $W$ with payments and losing bids $L$.

\begin{algorithm}[t!]
\caption{Greedy Value-Density Auction}
\label{alg:greedy_auction_compact}
\textbf{Input:} Bids, $\mathcal{B} = \{b_i = (\text{id}_i, \text{size}_i, \text{value}_i)\}$, $C$, $T_{\text{slot}}$ \\
\textbf{Output:} Winning bids $W$ (with payments $p_i$), Losing bids $L$

\begin{algorithmic}[1]
\State $\mathcal{E} \leftarrow \emptyset$ \Comment{Eligible bids with calculated properties}
\ForAll{$b_i = (\text{id}_i, s_i, v_i) \in \mathcal{B}$}
    \State $\text{bw\_req}_i \leftarrow s_i / (T_{\text{slot}} \times 3600)$ \Comment{Bandwidth required}
    \If{$v_i > 0$ and $\text{bw\_req}_i > 0$ and $\text{bw\_req}_i \leq C$}
        \State $\text{density}_i \leftarrow v_i / \text{bw\_req}_i$
        \State Add $(\text{id}_i, v_i, \text{bw\_req}_i, \text{density}_i)$ to $\mathcal{E}$
    \EndIf
\EndFor
\State Sort $\mathcal{E}$ in descending order of $\text{density}_i$
\State $W_{\text{temp}} \leftarrow \emptyset$, $C_{\text{used}} \leftarrow 0$
\ForAll{$e_j = (\text{id}_j, v_j, \text{bw\_req}_j, \text{density}_j) \in \mathcal{E}$}
    \If{$C_{\text{used}} + \text{bw\_req}_j \leq C$}
        \State Add $e_j$ to $W_{\text{temp}}$; $C_{\text{used}} \leftarrow C_{\text{used}} + \text{bw\_req}_j$
    \EndIf
\EndFor
\State $d^* \leftarrow \textsc{DetermineCriticalDensity}(\mathcal{E}, W_{\text{temp}})$
\State $W \leftarrow \emptyset$
\ForAll{$(\text{id}_k, v_k, \text{bw\_req}_k, \_) \in W_{\text{temp}}$}
    \State $p_k \leftarrow d^* \times \text{bw\_req}_k$ \Comment{Calculate payment}
    \State Add original bid $b_k$ corresponding to $\text{id}_k$ with payment $p_k$ to $W$
\EndFor
\State $L \leftarrow \mathcal{B} \setminus \{\text{bid part of } w \mid w \in W\}$
\State \Return{$W, L$}
\end{algorithmic}
\end{algorithm}

\begin{algorithm}[t!]
\caption{VCG Knapsack Auction}
\label{alg:vcg_knapsack_compact}
\textbf{Input:} Set of Bids $\mathcal{B}=\{b_i = (\text{id}_i, \text{size}_i, \text{value}_i)\}$, Capacity $C$, Slot Duration $T_{\text{slot}}$, Scale Factor $\alpha$ \\
\textbf{Output:} Set of Winning Bids $W$ (with payments $p_i$), Set of Losing Bids $L$

\begin{algorithmic}[1]
\State $C' \leftarrow \left\lceil C \times \alpha \right\rceil$ \Comment{Integer capacity}
\State $\mathcal{I} \leftarrow \emptyset$ \Comment{Eligible items for knapsack}
\ForAll{$b_i = (\text{id}_i, s_i, v_i) \in \mathcal{B}$}
    \State $\text{bw\_req}_i \leftarrow s_i / (T_{\text{slot}} \times 3600)$
    \State $w'_i \leftarrow \left\lceil \text{bw\_req}_i \times \alpha \right\rceil$ \Comment{Integer weight}
    \If{$v_i > 0$ and $w'_i > 0$ and $w'_i \leq C'$}
        \State Add $(\text{id}_i, v_i, w'_i, \text{original\_bid\_reference}_i)$ to $\mathcal{I}$
    \EndIf
\EndFor
\State $(SW_{\text{all}}, W_{\text{items}}) \leftarrow \textsc{SolveKnapsack}(\mathcal{I}, C')$ \Comment{Optimal value and winning items}
\State $W \leftarrow \emptyset$
\ForAll{$(\text{id}_j, v_j, w'_j, b_j^{\text{orig}}) \in W_{\text{items}}$}
    \State $(SW_{-j}, \_) \leftarrow \textsc{SolveKnapsack}(\mathcal{I} \setminus \{(\text{id}_j, v_j, w'_j, b_j^{\text{orig}})\}, C')$
    \State $p_j \leftarrow \max(0, SW_{-j} - (SW_{\text{all}} - v_j))$ \Comment{Calculate VCG payment}
    \State Add $(b_j^{\text{orig}}, p_j)$ to $W$
\EndFor
\State $W_{\text{ids}} \leftarrow \{\text{id}_j \mid (\text{id}_j, \_, \_, \_) \in W_{\text{items}}\}$
\State $L \leftarrow \{ b_i \in \mathcal{B} \mid \text{id}_i \notin W_{\text{ids}} \}$
\State \Return{$W, L$}
\end{algorithmic}
\end{algorithm}

\section{Evaluation}
\subsection{Methodology}

We evaluate the proposed mechanisms under diverse yet controlled conditions while maintaining realism. We employed a synthetic workload generation strategy grounded in the characteristics of real-world data transfers. Direct use of historical traces presents limitations, such as the inability to systematically vary load levels or network capacity and the absence of explicit user value information crucial for auction evaluation. Our synthetic approach addresses these limitations, enabling controlled, reproducible experiments. We first analyzed a large dataset of GridFTP transfer logs~\cite{alcfGridFTP} collected over a decade (2014–2024) from operational systems to characterize key statistical properties relevant to network scheduling. This characterization included extracting the empirical distribution of aggregate hourly transfer request arrivals to capture inherent burstiness, deriving separate empirical probability distributions for the sizes of PUT (upload) and GET (download) transfers, and determining the relative frequency of these operations.

Based on these statistics, synthetic workload traces were generated for a 7-day period. For each hour, the number of transfer requests was sampled with replacement from the empirical arrival distribution, and request times within the hour were assigned uniformly at random. Each request was labeled as PUT or GET according to the observed frequency observed frequency ($\approx 35\%$ PUT), with file sizes sampled from the corresponding empirical distribution. Because true user utility is unobservable, each request was assigned a synthetic value composed of a log-normal base component, a small size-proportional term, and random noise to ensure positive and heterogeneous values. To study contention effects, a load multiplier ($0.8\times$–$2.0\times$) scaled hourly arrivals while preserving the underlying arrival patterns.

We evaluated computational overhead on a CPU workstation (Intel i7-13700, 64\,GB RAM). For a 7-day synthetic workload with thousands of transfer requests, the full simulation completes within minutes, implying that solving the Winner Determination Problem for a single hourly slot requires only milliseconds to seconds. The Greedy Value Density Auction is faster due to its $O(N \log N)$ complexity, while the VCG Knapsack Auction also converges well within operational limits without GPU acceleration. Both mechanisms therefore introduce negligible overhead relative to FCFS baselines.

\subsection{Baseline Mechanisms for Comparison}

To contextualize the performance of our proposed auction-based scheduling approaches (Greedy Value-Density and VCG), we rigorously compare them against two standard, value-oblivious baselines. The first, \textit{First-Come, First-Served with Admission Control} (FCFS-AC), models a resource reservation strategy in which transfers are admitted only if their estimated bandwidth requirement can be met; otherwise, they are placed in a FIFO queue. This approach prioritizes predictability for admitted transfers but may introduce queuing delays. The second, \textit{First-Come, First-Served with Congestion} (FCFS-Congestion), represents a work-conserving system where all requests are admitted and the available bandwidth is dynamically shared proportionally among active transfers, resulting in variable completion times under load.

These FCFS variants exemplify common yet distinct strategies for managing resource contention without incorporating value considerations. They serve as comprehensive baselines for evaluating the benefits of our value-aware mechanisms, particularly in terms of queuing behavior, preemption dynamics, resource pricing, and congestion management.

\section{Results and Discussion}

This section evaluates the proposed auction mechanisms against FCFS baselines through simulation experiments. The figures compare four scheduling approaches: \textbf{Greedy} (our proposed Greedy Value-Density Auction), \textbf{VCG} (our proposed VCG Knapsack Auction), \textbf{FCFS (A)} (baseline First-Come, First-Served with Admission Control), and \textbf{FCFS (C)} (baseline First-Come, First-Served with Congestion/bandwidth sharing). Unless stated otherwise, results show performance across increasing load multipliers and varying network bandwidth capacities (faceted in plots).

\begin{figure}[t!] 
    \centering
    \includegraphics[width=0.9\linewidth]{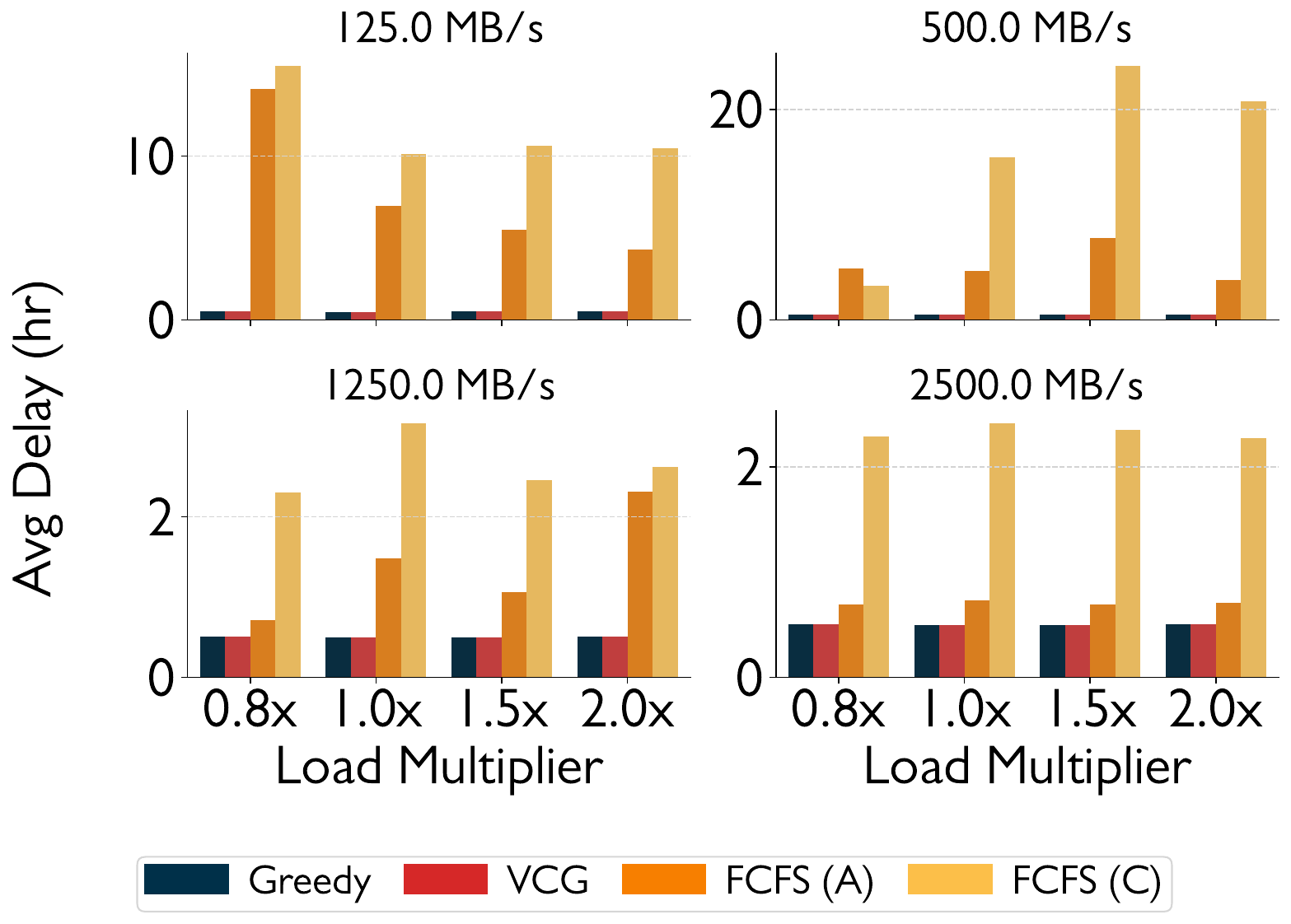} 
    \caption{Average completion delay (hr) across varying bandwidths and load multipliers. }
    \label{fig:avg_delay_individual}
\end{figure}

\begin{figure}[t!]
    \centering
    \includegraphics[width=0.9\linewidth]{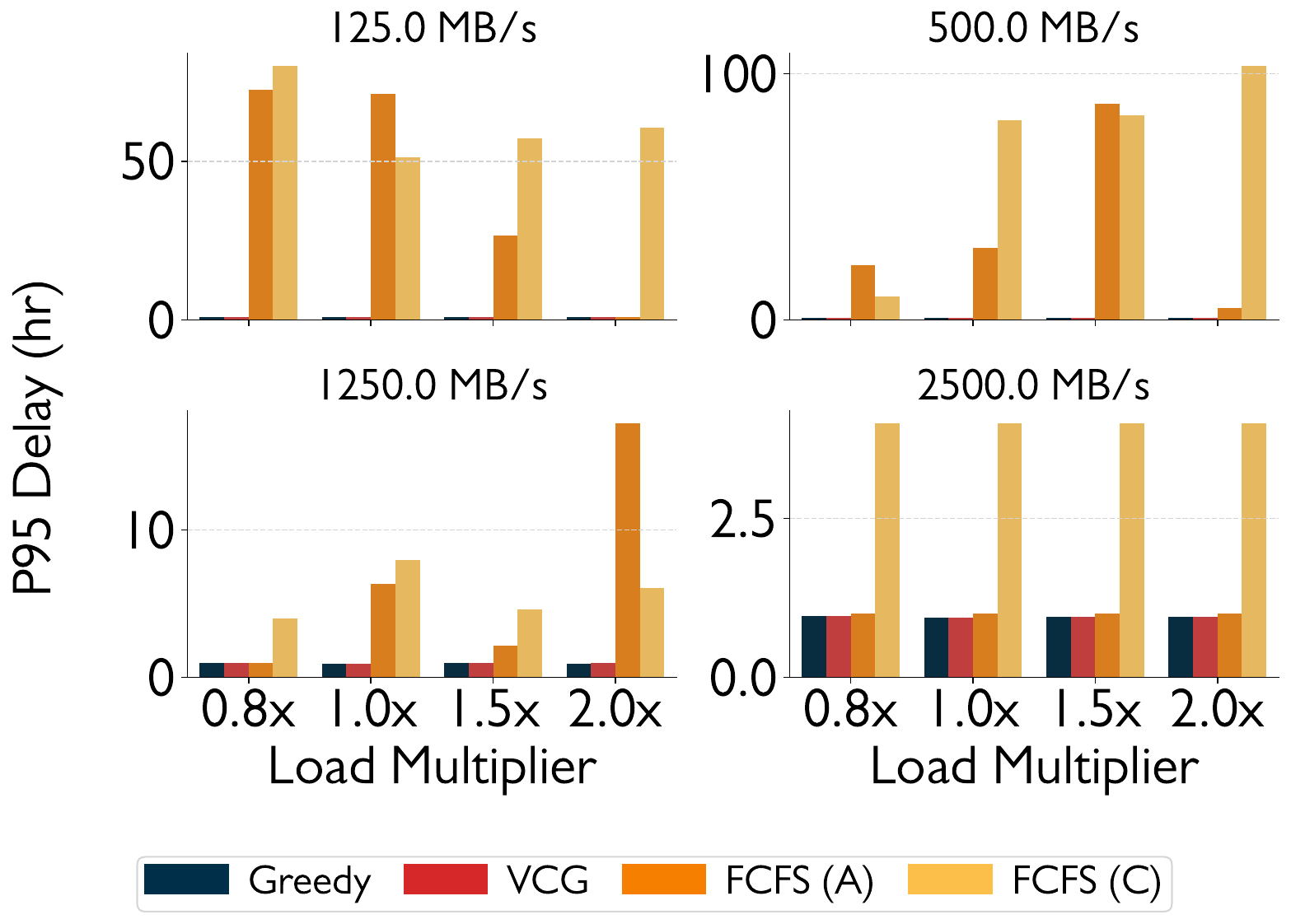} 
    \caption{95th Percentile (P95) completion delay (hr) across varying bandwidths and load multipliers.}
    \label{fig:p95_delay_individual}
\end{figure}

\begin{figure}[t!]
    \centering
    \includegraphics[width=0.8\linewidth]{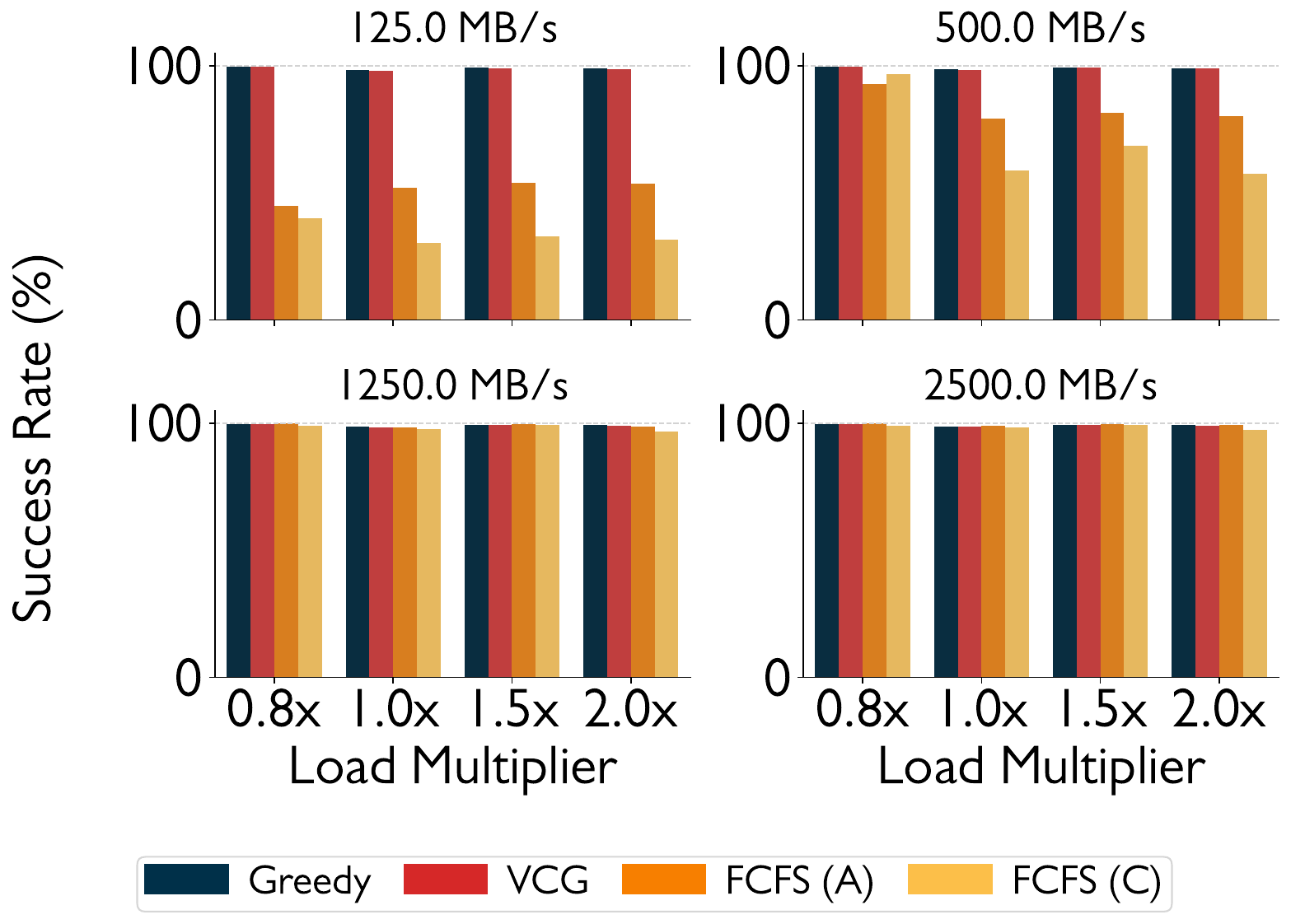} 
    \caption{Transfer success rate (\%) across varying bandwidths and load multipliers.}
    \label{fig:success_rate_individual} 
\end{figure}

\begin{figure}[t!]
    \centering
    \includegraphics[width=0.8\linewidth]{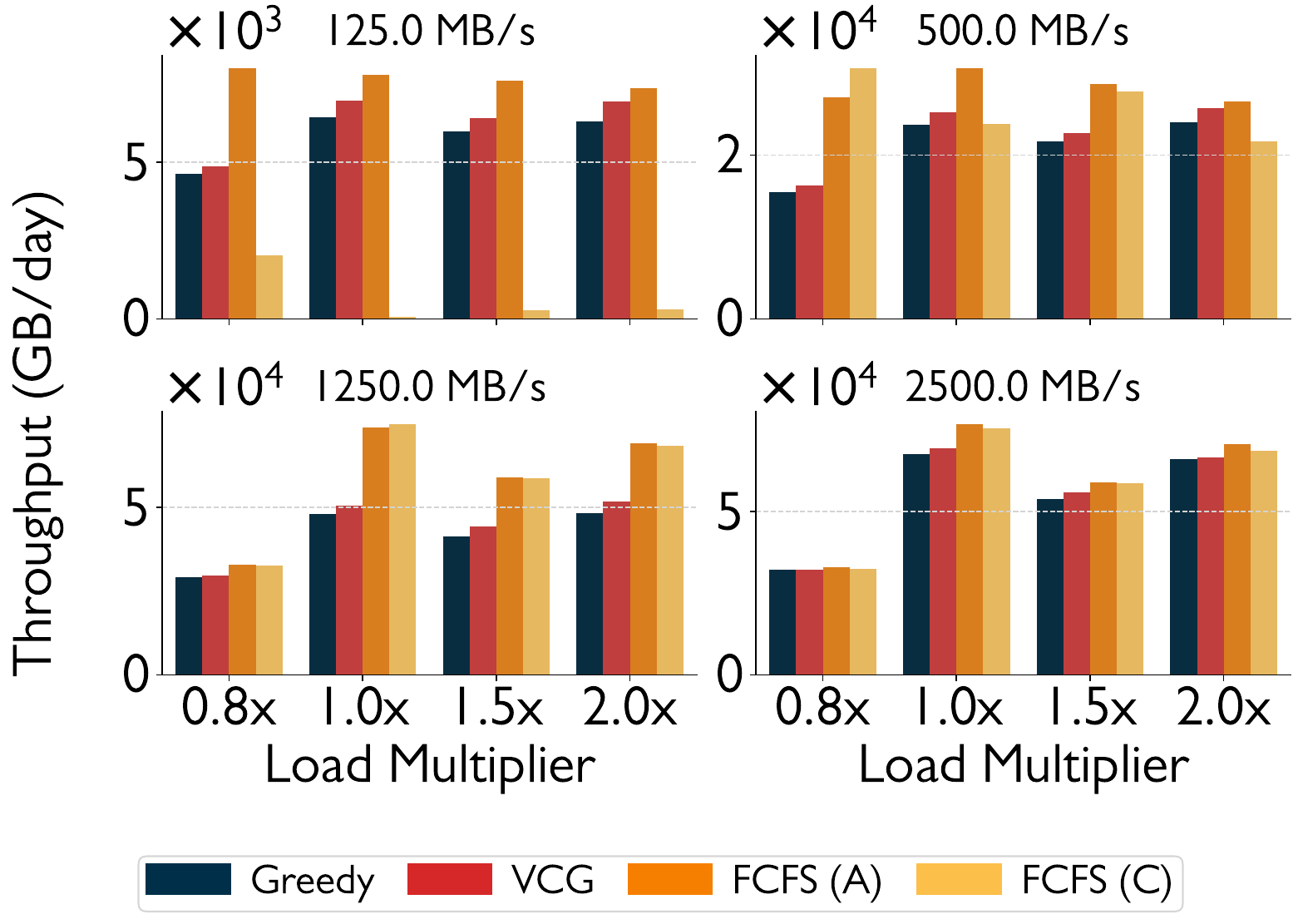} 
    \caption{System throughput (GB/day) across varying bandwidths and load multipliers. }
    \label{fig:throughput_individual} 
\end{figure}


\begin{figure}[t!]
    \centering
    \includegraphics[width=0.9\linewidth]{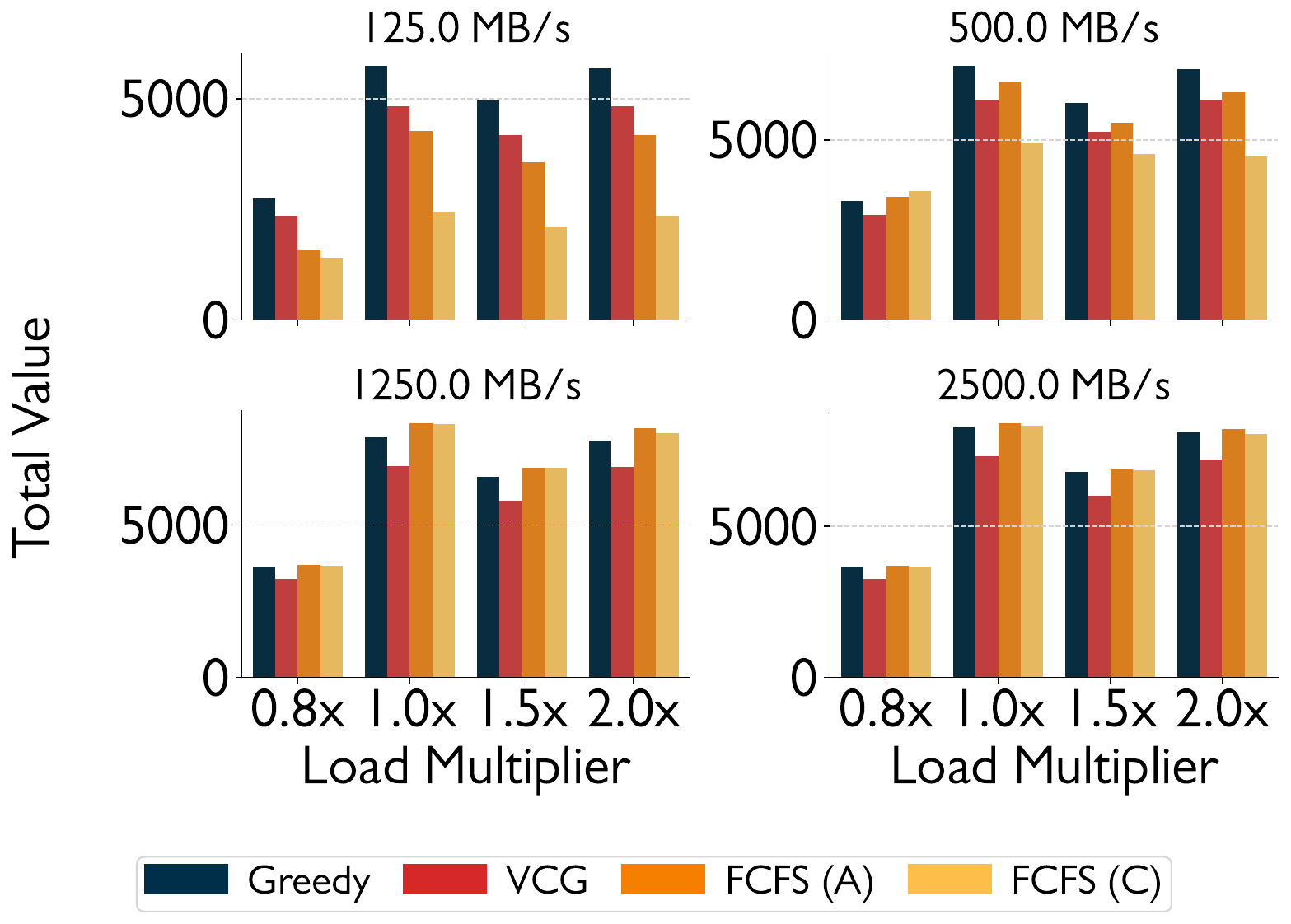} 
    \caption{Total value of completed transfers across varying bandwidths and load multipliers.}
    \label{fig:total_value_individual} 
\end{figure}

\begin{figure}[t!]
    \centering
    \includegraphics[width=0.9\linewidth]{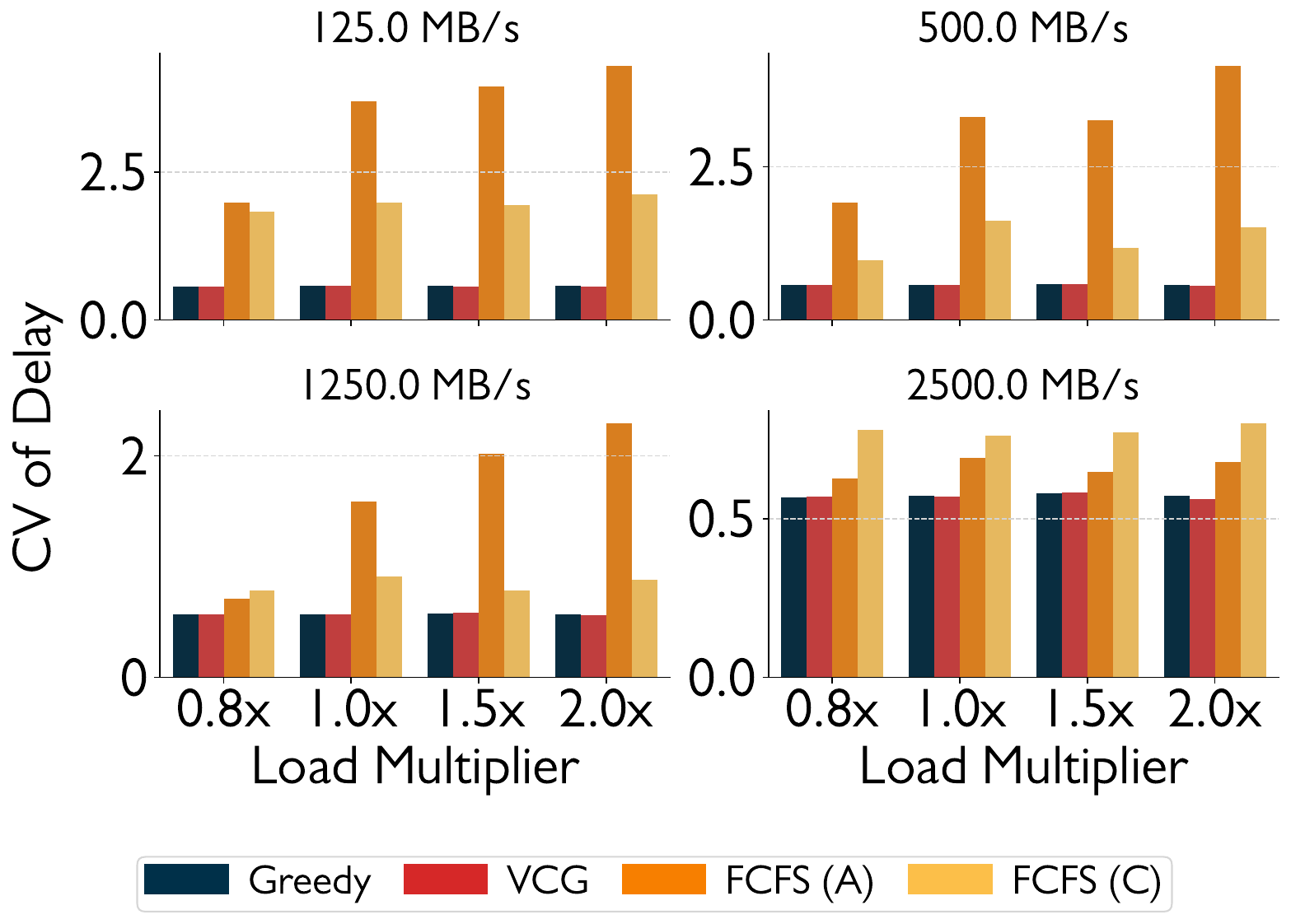} 
    \caption{Coefficient of Variation (CV) of Delay across varying bandwidths and load multipliers.}
    \label{fig:cv_delay_individual} 
\end{figure}

\subsection{Transfer Completion Delay}
\label{subsec:results_delay}

Figs.~\ref{fig:avg_delay_individual} and~\ref{fig:p95_delay_individual} present the performance of the evaluated mechanisms in terms of transfer completion delay, specifically focusing on the average delay (Figure ~\ref{fig:avg_delay_individual}) and the 95th percentile (P95) delay (Figure ~\ref{fig:p95_delay_individual}) for successfully completed transfers. Completion delay is a critical user-facing metric reflecting system responsiveness. A clear distinction emerges between the value-aware auction mechanisms (Greedy, VCG) and the value-oblivious FCFS baselines (FCFS (A), FCFS (C)). Both Greedy auction and VCG auction consistently achieve significantly lower average and P95 completion delays across all tested bandwidths and load levels compared to FCFS (A) and FCFS (C), as shown in Figs.~\ref{fig:avg_delay_individual} and~\ref{fig:p95_delay_individual}. The performance advantage of the auction mechanisms becomes particularly pronounced under higher load conditions (1.5x, 2.0x) and lower bandwidth capacities (125~MB/s, 500~MB/s), where the FCFS variants exhibit substantial delays, often exceeding several hours on average and tens of hours for the P95 metric (note the different y-axis scales across facets in these figures). This indicates that the auction mechanisms' ability to prioritize transfers based on value (or value density) effectively mitigates congestion and prevents the system from becoming overwhelmed, leading to much faster completions for the transfers that are admitted.

Interestingly, the delay performance difference between the Greedy Auction and the VCG auction is minimal across the tested scenarios when considering only completion time. Comparing the FCFS baselines, FCFS (C) generally results in higher average delays due to bandwidth sharing slowdown (Figure ~\ref{fig:avg_delay_individual}), while FCFS (A), despite potentially lower average delays in some light load scenarios, can suffer from extremely high P95 delays (e.g., $>$75 hours at 125~MB/s, 1.0x load in Figure ~\ref{fig:p95_delay_individual}), likely due to head-of-line blocking effects in its admission queue, especially at lower bandwidths.

\begin{tcolorbox}[colback=green!5!white, sharp corners]
\textbf{Finding 1:} \textit{Value-aware auction mechanisms (Greedy and VCG) drastically reduce both average and tail completion delays compared to value-oblivious FCFS approaches, particularly under network contention, as shown in Figs.~\ref{fig:avg_delay_individual} and~\ref{fig:p95_delay_individual}.}
\end{tcolorbox}

\subsection{Success Rate and System Throughput}
\label{subsec:results_success_throughput}

We evaluate transfer reliability and overall data volume via success rate (percentage of requests completed, Figure~\ref{fig:success_rate_individual}) and system throughput (GB/day, Figure~\ref{fig:throughput_individual}). Figure~\ref{fig:success_rate_individual} shows that auction mechanisms (Greedy and VCG) maintain near 100\% success rates across most scenarios, highlighting their reliability. In contrast, FCFS mechanisms, particularly FCFS (C), exhibit significantly degraded success rates under increasing load and lower bandwidths due to transfer slowdowns or prolonged queuing. At higher bandwidths, all mechanisms achieve high success rates. System throughput (Figure~\ref{fig:throughput_individual}) scales with bandwidth and generally increases with load for all approaches, though saturation occurs at high load. While FCFS (A) sometimes achieves the highest raw throughput by aggressively filling bandwidth, it does so at the cost of reliability and delay (Figures~\ref{fig:avg_delay_individual},\ref{fig:p95_delay_individual}, and\ref{fig:success_rate_individual}). Auction mechanisms closely match this throughput, with FCFS (C) consistently performing worst, especially at lower bandwidths.

\begin{tcolorbox}[colback=green!5!white, sharp corners]
\textbf{Finding 2:} \textit{Auction mechanisms ensure high transfer reliability (near 100\% success rate) even under significant load (Figure ~\ref{fig:success_rate_individual}), whereas FCFS mechanisms sacrifice success rate as contention increases. While FCFS (A) can maximize raw throughput, it does so at the cost of reliability and delay; auction mechanisms offer a better balance, achieving high throughput (Figure ~\ref{fig:throughput_individual}) alongside superior delay and success rate performance.}
\end{tcolorbox}

\subsection{Economic Efficiency and System Predictability}
\label{subsec:results_economic_predictability}


\begin{figure}[t!] 
    \centering
    \includegraphics[width=0.9\linewidth]{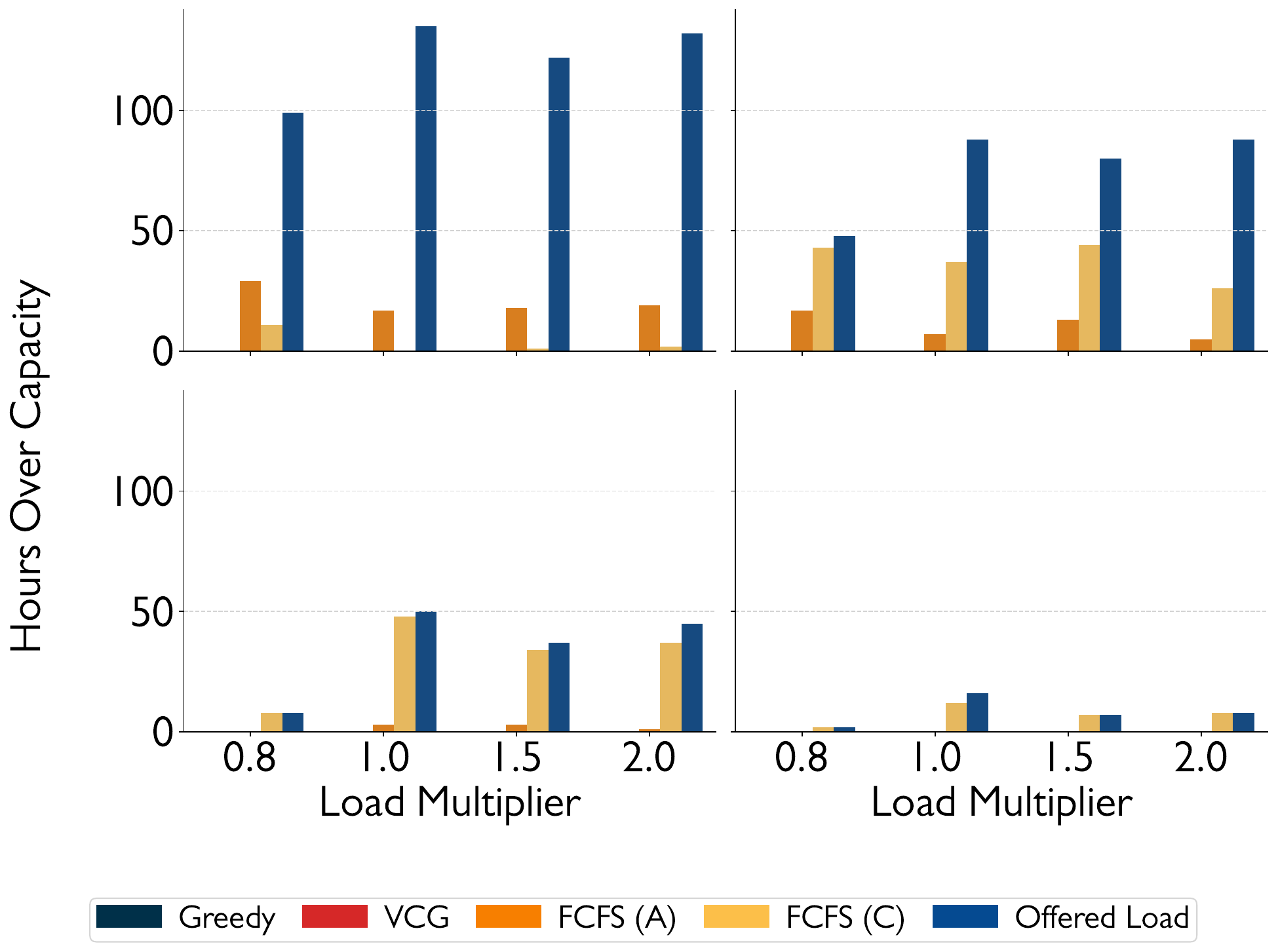} 
    \vspace{-3mm} 
    \caption{Hours where offered load and consumed load (by each mechanism) exceeded nominal link capacity, across varying bandwidths and load multipliers.}
    \label{fig:capacity_violations_individual}
\end{figure}
Beyond aggregate throughput, we evaluate the mechanisms based on their economic efficiency—measured by the total synthetic value of completed transfers—and their ability to provide predictable service delivery, particularly under varying load conditions. Figure \ref{fig:total_value_individual} illustrates the total value achieved by each mechanism across different loads and bandwidths. Complementing this, Figure \ref{fig:cv_delay_individual} (assuming this is the new reference for the CV of Delay plot) showcases the Coefficient of Variation (CV) of completion delays, offering insights into the predictability of each mechanism.
Figure ~\ref{fig:total_value_individual} shows that both the Greedy and VCG auctions consistently realize higher total value compared to the FCFS baselines, especially as load increases. While FCFS mechanisms complete transfers, their value-oblivious nature means they do not prioritize high-value requests, leading to lower overall economic efficiency. As the load multiplier increases, the total value achieved by FCFS mechanisms tends to saturate or even decrease (particularly for FCFS (C) at lower bandwidths), whereas the auction mechanisms generally sustain or slightly increase the total value captured, demonstrating better resource allocation under contention. Comparing the two auction mechanisms, the Greedy auction often achieves slightly higher total value than VCG in these simulations. This might seem counter-intuitive given VCG's optimality property regarding reported bids, but can occur due to the synthetic value model used, the greedy heuristic potentially admitting slightly different sets of transfers, and the discrete nature of the simulation slots.
Figure ~\ref{fig:cv_delay_individual} reveals significant differences in the predictability of service times across the evaluated mechanisms. The CV of Delay, which measures the delay variability relative to the mean, is markedly lower for both Greedy and VCG auctions across all tested bandwidths and load levels, typically remaining around 0.5-0.6. This indicates that completion delays under auction mechanisms are more tightly clustered around their average, signifying more predictable performance. In contrast, FCFS (A) and FCFS (C) exhibit substantially higher CV values, often ranging from 2.0 to over 4.0, particularly at lower bandwidths (125-1250 MB/s). This high relative variability implies that while some transfers might complete near the average delay, a significant portion can experience delays that are drastically different, making the service less predictable under these FCFS schemes. Even as overall system capacity increases (e.g., at 2500 MB/s), where all mechanisms show improved (lower) CV, the auction mechanisms maintain their advantage or perform comparably in terms of predictability.
\begin{tcolorbox}[colback=green!5!white, sharp corners]
\textbf{Finding 3:} \textit{Auction-based mechanisms achieve superior economic efficiency (higher total value completed, Figure \ref{fig:total_value_individual}) compared to FCFS baselines by prioritizing high-value transfers, especially under resource contention. Furthermore, auctions offer significantly more predictable service delivery, exhibiting substantially lower variability in completion delays (CV of Delay, Figure \ref{fig:cv_delay_individual}) than FCFS mechanisms across various load and bandwidth conditions.}
\end{tcolorbox}

\subsection{Load Dynamics and Utilization}

\begin{figure}[t!] 
    \centering
    \includegraphics[width=0.9\linewidth]{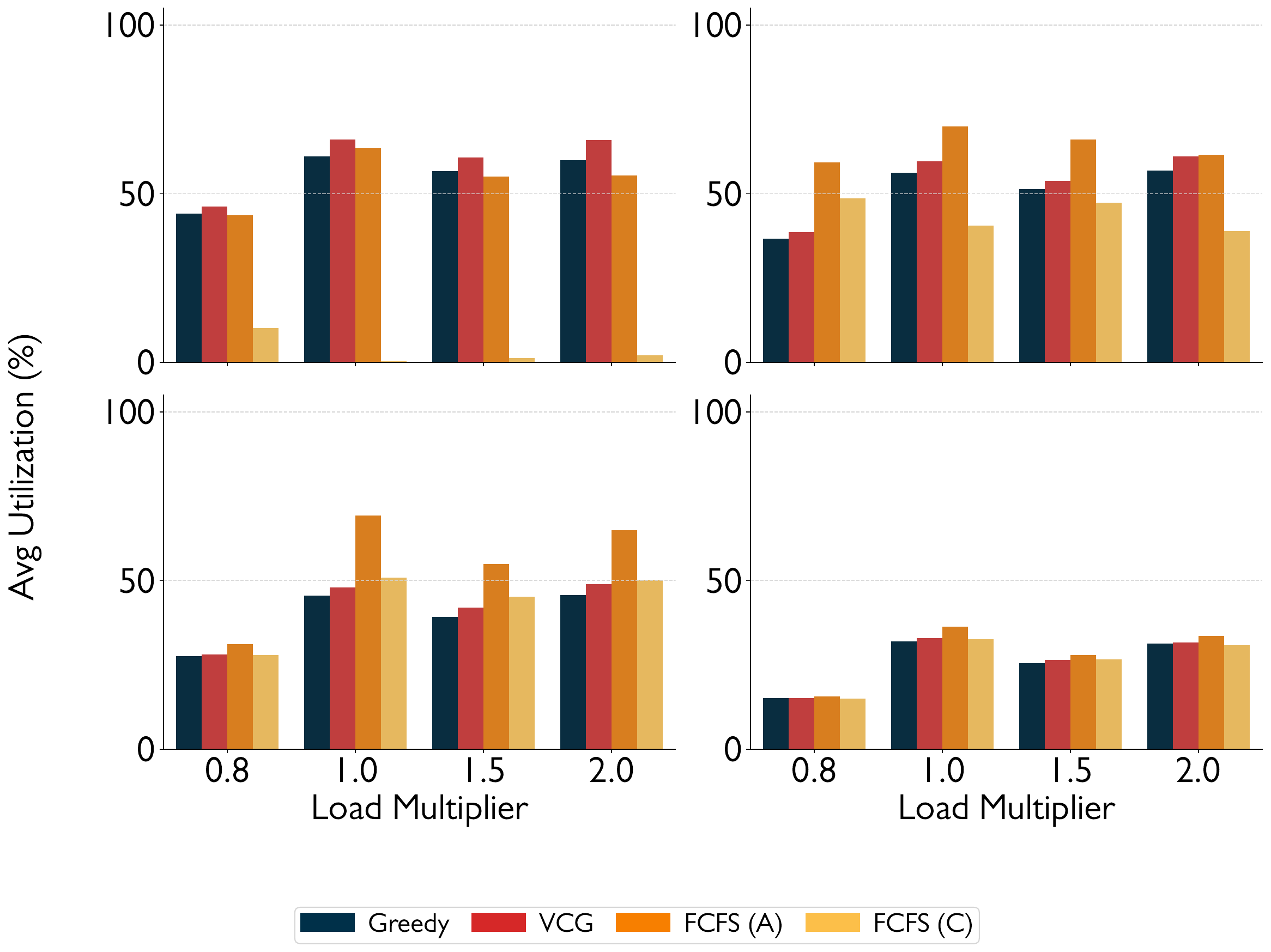} 
    \vspace{-3mm} 
    \caption{Average Network Utilization (\%) achieved by each mechanism across varying bandwidths and load multipliers. }
    \label{fig:avg_utilization_individual}
\end{figure}

\begin{figure}[t!] 
    \centering
    \includegraphics[width=0.9\linewidth]{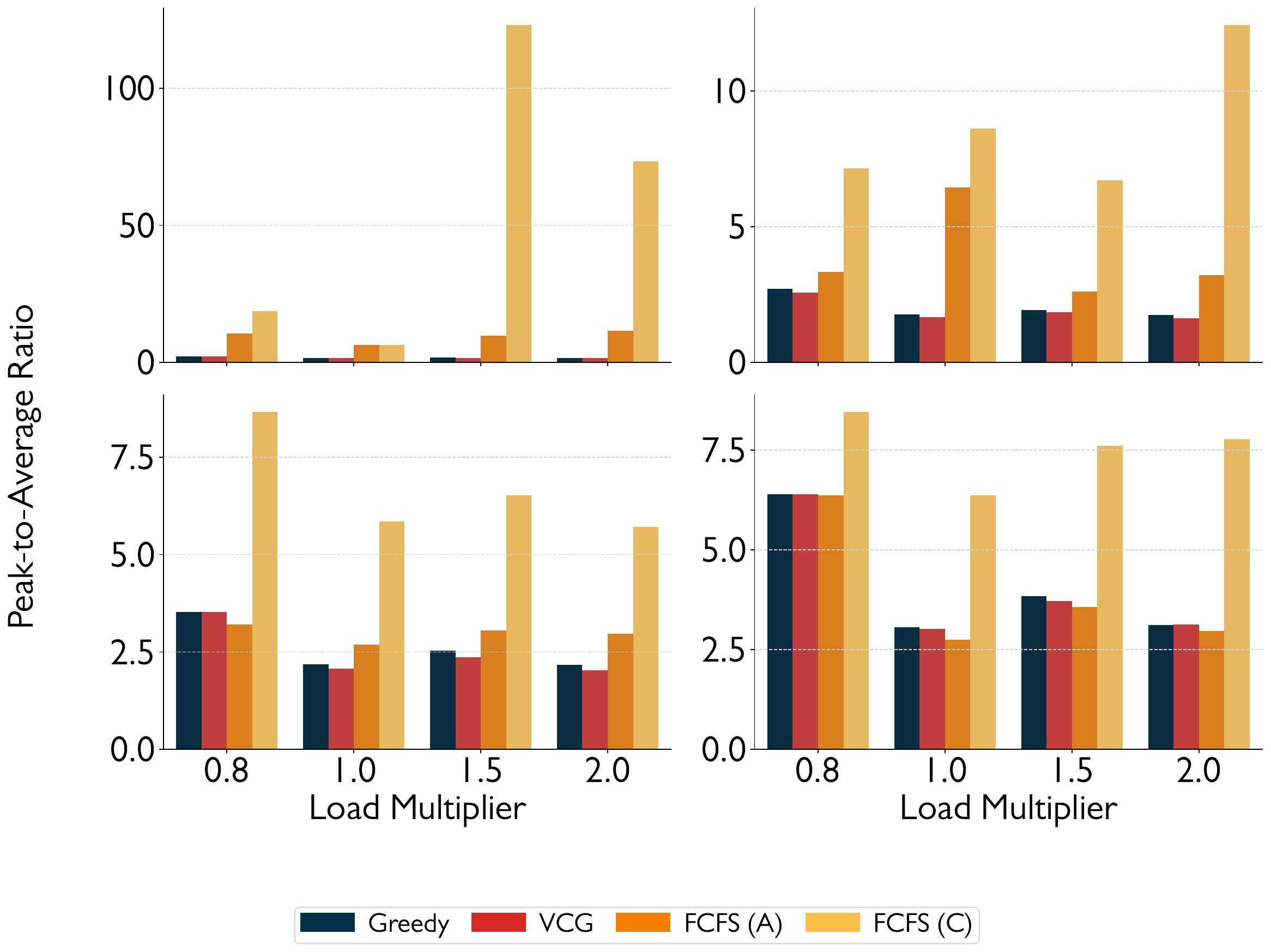} 
    \vspace{-3mm} 
    \caption{Load Volatility, measured by Peak-to-Average Ratio (PAR) of hourly consumed load, for each mechanism across varying bandwidths and load multipliers.}
    \label{fig:volatility_par_individual}
\end{figure}

We next examine network load management and capacity utilization through capacity violation frequency (Figure~\ref{fig:capacity_violations_individual}), average network utilization (Figure~\ref{fig:avg_utilization_individual}), and load volatility via Peak-to-Average Ratio (PAR, Figure~\ref{fig:volatility_par_individual}). Figure~\ref{fig:capacity_violations_individual} shows that while offered load frequently exceeds nominal capacity, especially at higher load multipliers, auction mechanisms (Greedy, VCG) and FCFS (A) effectively manage capacity with virtually no violations, unlike FCFS (C) which shows some overages likely due to transient bursts. Average network utilization (Figure~\ref{fig:avg_utilization_individual}) increases with offered load for all mechanisms, reaching 50-70\% under moderate to high load, with no single mechanism consistently dominating; FCFS (A) sometimes achieves slightly higher utilization by aggressively filling capacity, albeit at the cost of higher delays (Figures~\ref{fig:avg_delay_individual} and \ref{fig:p95_delay_individual}). Load stability, depicted by PAR in Figure~\ref{fig:volatility_par_individual}, reveals that FCFS (C) and, to a lesser extent, FCFS (A) exhibit higher volatility. In contrast, both auction mechanisms demonstrate substantially lower PAR, indicating more predictable and stable network usage due to their periodic, aggregate demand-based allocation.

\begin{tcolorbox}[colback=green!5!white, sharp corners]
\textbf{Finding 4:} \textit{Value-aware auctions and FCFS (A) effectively enforce network capacity limits (Figure~\ref{fig:capacity_violations_individual}). While all mechanisms achieve comparable average utilization under load (Figure~\ref{fig:avg_utilization_individual}), the auction mechanisms significantly reduce load volatility (lower PAR, Figure~\ref{fig:volatility_par_individual}), resulting in more stable network usage compared to FCFS baselines.}
\end{tcolorbox}


\section{Related Work}

\textbf{Infrastructure and Protocol-Level Solutions:} Significant work focuses on improving raw data transfer capability, including deploying high-performance Data Transfer Nodes (DTNs) within Science DMZs \cite{Dart2021}, developing adaptive protocols that dynamically tune parameters like concurrency and parallelism \cite{Arslan2018}, and mitigating network load through data reduction via summarization \cite{Foster2017} or lossy compression \cite{Li2021}. While these advancements maximize individual transfer performance, they do not solve the resource allocation problem when aggregate demand from competing users exceeds available capacity.

\textbf{Workflow Management and Co-Scheduling:} Higher-level solutions orchestrate complex scientific tasks through automation frameworks like Gladier \cite{gladier2024} and facility-specific APIs \cite{enders2020cross}, enabling event-driven workflows that adapt to system state \cite{Enders2024}. Research has explored co-scheduling network bandwidth alongside compute jobs \cite{enders2020cross}, and systems like CATCH use adaptive routing based on real-time performance \cite{Monti2011}. However, these approaches treat data movement via simple policies like First-Come, First-Served and lack mechanisms to arbitrate between concurrent transfers based on relative scientific value during high congestion.

\textbf{Value-Based and Economic Scheduling:} The most conceptually similar works apply economic principles for resource allocation. Value-based scheduling has been explored for allocating oversubscribed compute resources to maximize completed job value on power-constrained systems \cite{Kumbhare2017}, and wide-area flow scheduling has prioritized transfers based on size or urgency \cite{Giannakou2024}. However, these models target different resource characteristics—compute nodes are discrete and non-preemptive, whereas network bandwidth is divisible and shareable—and are not tailored to allocating shared ingress bandwidth among heterogeneous scientific data streams at HPC centers.

Our work bridges this gap by introducing an auction-based mechanism specifically designed for dynamic allocation of HPC ingress bandwidth guided by user-declared scientific value. By formulating the allocation as a multi-unit knapsack auction, we provide a method to resolve contention, improve economic efficiency, and prioritize bandwidth for the most impactful science—a dimension largely unaddressed by existing data movement tools and schedulers.

\section{Conclusion and Future Work}
HPC data ingestion is hampered by bandwidth bottlenecks and the limits of traditional methods like FCFS. This paper proposed and evaluated two value-driven auction mechanisms, Greedy Value Density and VCG Knapsack, which dynamically allocate ingestion bandwidth. Simulations confirmed auctions significantly outperform FCFS, reducing transfer completion delays by over 80\% in contended conditions and capturing over 100\% more total scientific value under high load, while also enhancing service predictability and transfer success rates. A key trade-off emerged: the Greedy auction offers lower computational overhead, while VCG provides stronger truthfulness incentives at a higher computational cost. Value-driven auctions offer a promising paradigm for faster data turnaround, improved system efficiency, and fairer resource access in HPC. Future work will focus on validating these mechanisms across other diverse HPC facilities to ensure generalizability, and developing hybrid approaches—such as utilizing the Greedy algorithm for rapid baseline allocation and dynamically triggering VCG during peak contention—prior to operational deployment.

\section{Acknowledgments}
This work was supported in part by the NSF under grant CNS-2300124.

\bibliographystyle{ieeetr}
\bibliography{ref}

\end{document}